\documentclass[conference]{IEEETran}
\usepackage[utf8]{inputenc}
\usepackage{hyperref}
\usepackage[numbered]{bookmark}
\usepackage{soul}
\usepackage{booktabs}
\usepackage{multirow}
\usepackage{float}
\usepackage{url}
\usepackage[small,it]{caption}
\usepackage{tcolorbox}
\usepackage{amsmath,amssymb,amsfonts}
\usepackage{algorithmic}
\usepackage{graphicx}
\usepackage{textcomp}
\usepackage{xcolor}
\usepackage{tabularx} 
\usepackage{balance}
\usepackage{enumitem}
\usepackage[a-2b,mathxmp]{pdfx}[2019/3/11]

\newcommand{\RQA}{How accurate is the ensemble part-of-speech tagger at the individual word level and how does this compare to the independent taggers?}

\newcommand{\RQB}{How accurate is the ensemble part-of-speech tagger at the identifier level and how does this compare to the independent taggers?}

\newcommand{\RQC}{What are the most frequently mis-used part-of-speech tags and grammar patterns?}

\makeatletter
\newcommand{\linebreakand}{%
  \end{@IEEEauthorhalign}
  \hfill\mbox{}\par
  \mbox{}\hfill\begin{@IEEEauthorhalign}
}
\makeatother
\author{Christian D. Newman\IEEEauthorrefmark{1}, Michael J. Decker\IEEEauthorrefmark{2}, Reem S. Alsuhaibani\IEEEauthorrefmark{3}, Anthony Peruma\IEEEauthorrefmark{1}, Mohamed Wiem Mkaouer\IEEEauthorrefmark{1},\\ Satyajit Mohapatra\IEEEauthorrefmark{1}, Tejal Vishnoi\IEEEauthorrefmark{1}, Marcos Zampieri\IEEEauthorrefmark{4}, Timothy J. Sheldon\IEEEauthorrefmark{5}, Emily Hill\IEEEauthorrefmark{6}
\\
\\
\IEEEauthorrefmark{1}Software Engineering Deptartment, Rochester Institute of Techonology, Rochester, NY\\
\IEEEauthorrefmark{2}Software Engineering Department, Bowling Green State University, Bowling Green, OH\\
\IEEEauthorrefmark{3}Computer Science Department, Prince Sultan University, Riyadh, Saudi Arabia\\
\IEEEauthorrefmark{4}Language Technology Group, Rochester Institute of Techonology, Rochester, NY\\
\IEEEauthorrefmark{5}Financial Services Sector, Risk and Compliance, BNY Mellon, Pittburgh, PA\\
\IEEEauthorrefmark{6}Department of Math and Computer Science, Drew University, Madison, NJ}

\begin{document}

\title{An Ensemble Approach for Annotating Source Code Identifiers with Part-of-speech Tags} 
% \author{
%     \IEEEauthorblockN{Christian~D.~Newman\IEEEauthorrefmark{1}, Michael~J.~Decker\IEEEauthorrefmark{2}, Anthony~Peruma\IEEEauthorrefmark{1}, Reem~Alsuhaibani\IEEEauthorrefmark{3}, Mohamed~Wiem~Mkaouer\IEEEauthorrefmark{1},
%     Satyajit~Mohapatra\IEEEauthorrefmark{1}, Tejal~Vishoi\IEEEauthorrefmark{1},
%     Marcos~Zampieri\IEEEauthorrefmark{1}, Emily~Hil\IEEEauthorrefmark{4}}
    
%     \IEEEauthorblockA{\IEEEauthorrefmark{1}Rochester institute of Technology
%     \\cnewman@se.rit.edu, \{axp6201,mwmvse,sm6571,tv4466,mazgla\}@rit.edu}
    
%     \IEEEauthorblockA{\IEEEauthorrefmark{2}Bowling Green State University
%     \\\{mdecke\}@bgsu.edu}
    
%     \IEEEauthorblockA{\IEEEauthorrefmark{3}Kent State University
%     \\\{rahlsuhai\}@kent.edu}
    
%     \IEEEauthorblockA{\IEEEauthorrefmark{4}Drew University
%     \\\{emhill\}@drew.edu}
% } 
\maketitle
\thispagestyle{plain}
\pagestyle{plain}
\begin{abstract}
    This paper presents an ensemble part-of-speech tagging approach for source code identifiers. Ensemble tagging is a technique that uses machine-learning and the output from multiple part-of-speech taggers to annotate natural language text at a higher quality than the part-of-speech taggers are able to obtain independently. Our ensemble uses three state-of-the-art part-of-speech taggers: SWUM, POSSE, and Stanford. We study the quality of the ensemble's annotations on five different types of identifier names: function, class, attribute, parameter, and declaration statement at the level of both individual words and full identifier names. We also study and discuss the weaknesses of our tagger to promote the future amelioration of these problems through further research. Our results show that the ensemble achieves 75\% accuracy at the identifier level and 84-86\% accuracy at the word level. This is an increase of +17\% points at the identifier level from the closest independent part-of-speech tagger.
\end{abstract}
\section{Introduction}
\label{introduction}

Program comprehension is a significant factor in the time it takes to develop and maintain software \cite{Corbi1989, Martin:2008}. Developers spend much more time reading code than they spend writing; 10 times more by some estimates \cite{Martin:2008}. Increased understanding of developer comprehension will lead to approaches that not only augment the ability of developers and program analysis tools to be productive, but also improve the accessibility of software development (e.g., by supporting programmers that prefer top-down or bottom-up comprehension styles \cite{vonMayrhauser:1997,Fisher:2006}) and help developers avoid stress stemming from code that is hard to understand. One of the primary ways a developer comprehends code is by reading identifier names, which make up on average about 70\% of the characters found in a body of code \cite{Deissenbock:2005}. Therefore, improving identifier naming practices can have a significant, positive impact on comprehension. 

One challenge to studying identifiers is the difficulty in understanding how to map the meaning of natural language phrases to the behavior of the code. For example, when a developer names a method, the name should reflect the behavior of the method such that another developer can understand what the method does without the need to read the method body. Understanding this connection between name and behavior presents challenges for humans and tools; both of which use this relationship to comprehend, generate, or critique code. A second challenge lies in the natural language analysis techniques themselves, many of which are not trained to be applied to software \cite{Binkley:2018}; introducing significant threats \cite{Jongeling:2017}. Addressing these problems is vital to improving the developer experience and augmenting tools which leverage natural language.

Analysis of identifier names can be done in many ways, including word frequency analysis \cite{Manning:1999} (e.g., ngrams) or semantic analysis using lexical ontologies like wordnet \cite{miller1995wordnet}. These are applied to a large number of problems, including rename refactoring analysis \cite{Liu:2019, Arnaoudova:2014,Perumascam, PERUMA2020110704}, linguistic anti-patterns \cite{Arnaoudova:2013}, identifier splitting \cite{Hill:2014}, and part-of-speech tagging \cite{HillSWUM:2010, Gupta:2013, Olney2016}. In this paper, we focus on part-of-speech tagging (POS); a technique whereby words in a sentence, or in an identifier in this case, are annotated based on the role they play within the context of the words surrounding them or based on their typical usage in the case where we are dealing with a single-word identifier. Part-of-speech tagging is one of the most popular methods for measuring the natural language semantics of identifier names and has been used in numerous other research  \cite{Host:2009, Gupta:2013, Shepherd:2009, butler:2011, butler:2015,Liblit06cognitiveperspectives, Arnaoudova:2013, PERUMA2020110704,posit:2020}. Unfortunately, part-of-speech taggers for identifiers are still inaccurate \cite{Newman2020,Olney2016}, making it difficult to trust their output. 

The goal of this paper is to discuss and present an ensemble tagging technique that improves the accuracy of part-of-speech taggers, and supports a larger variety of POS tags than other software engineering based POS taggers, specifically SWUM and POSSE \cite{Gupta:2013, HillSWUM:2010}. The ensemble approach uses machine-learning algorithms such as Decision Tree \cite{decisiontrees} and Random Forest \cite{randomforest}, which are common in other software research tasks \cite{Roy2020ICPC,AlOmar2021ESWA} and have been used for part-of-speech tagging of standard English documents \cite{DellOrletta2009EnsembleSF}. The main contributions of this work are as follows:
\begin{enumerate}
    \item An implementation of the most accurate (to-date) part-of-speech tagger for source code identifiers, built using data that was curated via significant manual-annotation effort made by the authors in prior work \cite{Newman2020}. This approach is trained to support more types of annotations (i.e., POS tags) specifically oriented for source code than any other approach currently available. In addition, the ensemble has been made fully available (see Section \ref{methodology}), and is intended for long term support by the research team as we expand the training set and include identifiers from different contexts (e.g., test code).
    \item Confirmation of observations we made in prior work \cite{Newman2020} that indicate 1) the importance of the position of a word in an identifier, 2) the importance of the context of an identifier when annotating using part-of-speech, and 3) the complementarity of three part-of-speech taggers.
    \item An expanded set of manually-annotated identifiers, based on the original set constructed in \cite{Newman2020}, that can be used to train and create other tagging approaches or for other natural language problems. As with the implementation, this has been made fully available to the research community (see Section \ref{methodology}).
    \item A thorough evaluation of the ensemble approach at both the identifier- and word-level, including a discussion of the features, which were empirically derived by the authors in prior work \cite{Newman2020}, that most positively influence the tagger's performance.
    \item A discussion that provides a clear path for future work on part-of-speech tagger accuracy and effectiveness. This discussion is based on the authors' experience in prior work \cite{Newman2020, 2021:ICPC:METHODS} combined with phenomena observed during the evaluation of the tagger.
\end{enumerate}

In addition to advancing the state-of-the-art of part-of-speech tagging, we also discuss where our approach is still weak; highlighting situations to which it may not sufficiently generalize due to potential limitations in our technique and our dataset. We answer the following Research Questions (RQs):

\textbf{RQ1: \RQA}  
In prior research \cite{Newman2020}, we found that the output of three part-of-speech taggers complemented one another by applying them all to a manually-curated dataset of grammar patterns. This question will address just how much we can improve the accuracy of part-of-speech taggers on source code by combining the output of these taggers using machine learning. 

\textbf{RQ2: \RQB} In RQ1, we explore word-level accuracy. In RQ2, we will look at how accurate our ensemble is when it must annotate the entire identifier correctly, since, as shown in prior work \cite{Newman2020}, even if a tagger has high accuracy on individual words, it may have low accuracy on full identifiers.

\textbf{RQ3: \RQC} This question investigates whether there are patterns in the way our approach mis-annotates identifiers. We explore these cases and discuss what further information the ensemble requires in order to handle these cases properly. For example, in prior work \cite{Newman2020}, we found that implementation details have an effect on the correct tag sequence for certain identifiers and are thus required to properly tag these identifiers. In this question, we take a deeper look at this problem among others.

% \textcolor{red}{\textbf{RQ4: \RQD} The final question investigates how much each feature improved, or degraded, the performance of our best ensembles. The motivation for this question is to reveal what characteristics of identifiers are most important to deciding on a POS annotation, according to our approach, and discuss why those features work best. While this question will not universally answer what the best feature set for any tagger is, it will provide some insight into the relationship between some identifier characteristics and POS annotations.}
This paper is organized as follows: Section \ref{grammarpatterndef} provides the necessary definitions and background to understand the paper. Section \ref{sec:related} explains related work.  Our methodology is detailed in Section \ref{methodology}, and Section \ref{setup} explains the Evaluation Setup. Section \ref{evaluation} presents the evaluation of our ensemble and answers to our RQs. Section \ref{threats} elaborates on our threats to validity and Section \ref{Discussion+conclusion} summarizes our results, discusses future work, and concludes.

\section{Definitions \& Grammar Pattern Generation}
\label{grammarpatterndef}
The application of a part-of-speech tagger to an identifier results in a grammar pattern \cite{Newman2020}. A \textit{grammar pattern} is the sequence of part-of-speech tags (also referred to as annotations) assigned to individual words within an identifier. For example, for an identifier called GetUserToken, we assign a grammar pattern by splitting the identifier into its three constituent words: Get, User, and Token. We then run the split-sequence (i.e., Get User Token) through a part-of-speech tagger to get its grammar pattern: Verb Noun-adjunct Noun, which can help us understand how the individual words in this identifier are related. The advantage to using grammar patterns for identifier analysis is that a given pattern is not unique to any individual identifier, but is shared with many potential identifiers that use similar words. Thus, we can relate identifiers that contain different words to one another using their grammar pattern; GetUserToken, RunUserQuery, and WriteAccessToken share the same grammar pattern and, while they do not express the exact same semantics, there are similarities in their semantics which their grammar patterns reveal. Specifically, a verb (get, run, write) applied to a noun (token, query) with a specific role/context (user, access).

\begin{table}[]
\centering
\caption{Examples of grammar patterns}
\label{Table:GrammarPatternExample}
\begin{tabular}{@{}ll@{}}
\toprule
\multicolumn{1}{c}{\textbf{Identifier Example}} & \multicolumn{1}{c}{\textbf{Grammar  Pattern}} \\ \midrule
1. GList* \textbf{tile list head} = NULL; & adjective adjective noun \\
2. GList* \textbf{tile list tail} = NULL; & adjective adjective noun \\
3. Gulong \textbf{max tile size} = 0; & adjective adjective noun \\
4. GimpWireMessage \textbf{msg}; & noun \\
5. \textbf{g list remove link} (tile list head, list) & preamble noun verb noun \\
6. \textbf{g list last} (list) & preamble adjective noun \\
7. \textbf{g assert} (tile\_list\_head != tile\_list\_tail); & preamble verb \\
\bottomrule
\end{tabular}
\end{table}
\begin{table}[]
\centering
\caption{Part-of-speech categories in dataset and supported by ensemble}
\label{tab:posusedtable}
\begin{tabular}{@{}ccl@{}}
\toprule
\textbf{Abbreviation} & \textbf{Expanded Form} & \multicolumn{1}{c}{\textbf{Examples}} \\ \midrule
N & noun & Disneyland, shoe, faucet, mother \\ \midrule
DT & determiner & the, this, that, these, those, which \\ \midrule
CJ & conjunction & and, for, nor, but, or, yet, so \\ \midrule
P & preposition & behind, in front of, at, under, above \\ \midrule
NPL & noun plural & Streets, cities, cars, people, lists \\ \midrule
NM & \begin{tabular}[c]{@{}c@{}}noun modifier \\ (adjectives, \\ noun-adjuncts)\end{tabular} & red, cold, hot, scary, beautiful, small \\ \midrule
V & verb & Run, jump, spin \\ \midrule
VM & \begin{tabular}[c]{@{}c@{}}verb modifier \\ (adverb)\end{tabular} & Very, loudly, seriously, impatiently \\ \midrule
PR & pronoun & she, he, her, him, it,we,they,them \\ \midrule
D & digit & 1, 2, 10, 4.12, 0xAF \\ \midrule
PRE & preamble & Gimp, GLEW, GL, G, p, m, b \\ \bottomrule
\end{tabular}%
\end{table}
\subsection{Annotating identifiers}\label{annotations}
%The first step to annotating an identifier is ti split it into its constituent terms. We use Spiral for this purpose \cite{HuckaSpiral}. Spiral is an open-source tool combining several splitting approaches into one Python package. From this package, we used heuristic splitting and manually corrected mistakes made by the splitter, which we discuss in the next section. We generate grammar patterns by running each individual tagger on each identifier after applying our splitting function. Additionally, we give examples of grammar patterns in Table~\ref{Table:GrammarPatternExample}, which shows a set of identifiers on the left and the corresponding grammar pattern on the right.

Since part of the goal of this paper is to study an ensemble part-of-speech tagger, we use multiple taggers; POSSE \cite{Gupta:2013}, SWUM \cite{HillSWUM:2010}, and Stanford \cite{Toutanova:StanfordTagger}\footnote{Version: 3.9.2, taggermodel: english-bidirectional-distsim.tagger, jdk version: openJDK 11.0.7}. POSSE and SWUM are part-of-speech taggers created specifically to be run on software identifiers; they are trained to deal with the specialized context in which identifiers appear. Both POSSE and SWUM take advantage of static analysis to provide annotations. For example, they will look at the return type of a function to determine whether the word \textit{set} is a noun or a verb. Additionally, they are both aware of common naming structures in identifier names. For example, methods are more likely to contain a verb in certain positions within their name (e.g., at the beginning) \cite{Gupta:2013,HillSWUM:2010}. They leverage this information to help determine what part-of-speech to assign different words. Stanford is a popular part-of-speech tagger for general natural language (e.g., English) text. For our study, Stanford provides a baseline; it is not specialized for source code but was found to be reasonably accurate on method names \cite{Olney2016} and shown to be competitive with POSSE \cite{Newman2020}.

This paper uses the part-of-speech tags given in Table~\ref{tab:posusedtable}. Note, in this table, the \textit{preamble} category, which does not exist in general natural language part-of-speech tagging approaches but has been defined and discussed in prior work \cite{HillSWUM:2010, Newman2020}. We use the definition from \cite{Newman2020}: A Preamble is an abbreviation that occurs at the beginning of an identifier and does one of the following: 
\begin{enumerate}[nolistsep]
    \item Namespaces an identifier without augmenting the reader's understanding of its behavior (e.g., XML in XML\_Reader is not a preamble)
    \item Provides language-specific metadata about an identifier (e.g., identifies pointers or member variables)
    \item Highlights an identifier's type. When a preamble is highlighting an identifier's type, the type's inclusion must not add any new information to the identifier name.
\end{enumerate}
For example, given an identifier \textit{float* fPtr}, 'f' does not add any information about the identifier's role within the system, but reminds the developer that it has a type 'float'. However, given an identifier \textit{char* sPtr}, 's' informs the developer that this is a c-string as opposed to a pointer to some other type of character array; 's' would not be considered a preamble under this definition. Additionally, some developers will put \textit{p\_} in front of pointer variables or m\_ in front of variables that are members of a class; these are due to naming conventions and/or Hungarian notation \cite{butler2010exploring,butler:2015,hillamap, hungariannotation}. In the GIMP and GLEW open-source projects, GIMP and G\_ are preambles to many variables, as seen in the Gimp example in Table~\ref{Table:GrammarPatternExample}. Intuitively, the reason for identifying preambles in an identifier is because they do not provide any information with respect to the identifier's role within the system's domain. Instead, they provide one of the three types of information above. For this reason, when analyzing identifiers, tools should be able to determine when a word is a preamble so that they do not make false assumptions about the word's descriptive purpose.

Another tag to note from Table \ref{tab:posusedtable} is \textit{noun modifier (NM)}, which is annotated on words that could be considered either pure adjectives or noun-adjuncts. A noun-adjunct is a word that is typically a noun but is being used as an adjective. An example of this is the word \textit{bit} in the identifier \textit{bitSet}. In this case, \textit{bit} is a noun which describes the type of \textit{set}, i.e., it is a set of bits. So we consider it a noun-adjunct. These are found in English (e.g., compound words), but generally not annotated as their own individual part-of-speech tag. Refer to our prior work for more information \cite{Newman2020}.

\begin{table}[]
\centering
\caption{How Penn Treebank annotations were mapped to the reduced set of annotations}
\label{tab:penntreebankmap}
\resizebox{!}{.28\paperheight}{%
\begin{tabular}{@{}c|c@{}}
\toprule
{\color[HTML]{000000} \begin{tabular}[c]{@{}c@{}}Penn Treebank\\ Annotation\end{tabular}} & {\color[HTML]{000000} \begin{tabular}[c]{@{}c@{}}Annotation Used\\ In Study\end{tabular}} \\ \midrule
{\color[HTML]{000000} Conjunction (CC)} & {\color[HTML]{000000} Conjunction (CJ)} \\ \midrule
{\color[HTML]{000000} Digit (CD)} & {\color[HTML]{000000} Digit (D)} \\ \midrule
{\color[HTML]{000000} Determiner (DT)} & {\color[HTML]{000000} Determiner (DT)} \\ \midrule
{\color[HTML]{000000} Foreign Word (FW)} & {\color[HTML]{000000} Noun (N)} \\ \midrule
{\color[HTML]{000000} Preposition (IN)} & {\color[HTML]{000000} Preposition (P)} \\ \midrule
{\color[HTML]{000000} Adjective (JJ)} & {\color[HTML]{000000} Noun Modifier (NM)} \\ \midrule
{\color[HTML]{000000} Comparative Adjective (JJR)} & {\color[HTML]{000000} Noun Modifier (NM)} \\ \midrule
{\color[HTML]{000000} Superlative Adjective (JJS)} & {\color[HTML]{000000} Noun Modifier (NM)} \\ \midrule
{\color[HTML]{000000} List Item (LS)} & {\color[HTML]{000000} Noun (N)} \\ \midrule
{\color[HTML]{000000} Modal (MD)} & {\color[HTML]{000000} Verb (V)} \\ \midrule
{\color[HTML]{000000} Noun Singular (NN)} & {\color[HTML]{000000} Noun (N)} \\ \midrule
{\color[HTML]{000000} Proper Noun (NNP)} & {\color[HTML]{000000} Noun (N)} \\ \midrule
{\color[HTML]{000000} Proper Noun Plural (NNPS)} & {\color[HTML]{000000} Noun Plural (NPL)} \\ \midrule
{\color[HTML]{000000} Noun Plural (NNS)} & {\color[HTML]{000000} Noun Plural (NPL)} \\ \midrule
{\color[HTML]{000000} Personal Pronoun (PRP)} & {\color[HTML]{000000} Pronoun (PR)} \\ \midrule
{\color[HTML]{000000} POSSEssive Pronoun (PRP\$)} & {\color[HTML]{000000} Pronoun (PR)} \\ \midrule
{\color[HTML]{000000} Adverb (RB)} & {\color[HTML]{000000} Verb Modifier (VM)} \\ \midrule
{\color[HTML]{000000} Comparative Adverb (RBR)} & {\color[HTML]{000000} Verb Modifier (VM)} \\ \midrule
{\color[HTML]{000000} Particle (RP)} & {\color[HTML]{000000} Verb Modifier (VM)} \\ \midrule
{\color[HTML]{000000} Symbol (SYM)} & {\color[HTML]{000000} Noun (N)} \\ \midrule
{\color[HTML]{000000} To Preposition (TO)} & {\color[HTML]{000000} Preposition (P)} \\ \midrule
{\color[HTML]{000000} Verb (VB)} & {\color[HTML]{000000} Verb (V)} \\ \midrule
{\color[HTML]{000000} Verb (VBD)} & {\color[HTML]{000000} Verb or Noun Modifier (V or NM)} \\ \midrule
{\color[HTML]{000000} Verb (VBG)} & {\color[HTML]{000000} Verb or Noun Modifier (V or NM)} \\ \midrule
{\color[HTML]{000000} Verb (VBN)} & {\color[HTML]{000000} Verb or Noun Modifier (V or NM)} \\ \midrule
{\color[HTML]{000000} Verb (VBP)} & {\color[HTML]{000000} Verb (V)} \\ \midrule
{\color[HTML]{000000} Verb (VBZ)} & {\color[HTML]{000000} Verb (V)} \\ \bottomrule
\end{tabular}
}
\end{table}

\begin{figure}[h]
\centering
\includegraphics[]{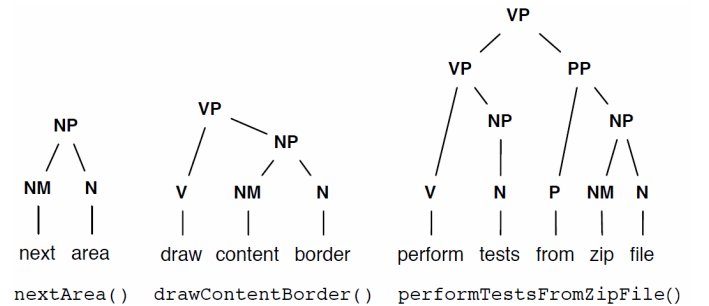}
\caption{Examples of noun, verb, and prepositional phrases}
\label{fig:example_phrasal_concepts}
\end{figure}

The tagset in Table~\ref{tab:posusedtable} is a smaller set than some standard natural language part-of-speech tagsets, such as the Penn Treebank tagset used by Stanford \cite{Toutanova:StanfordTagger}, due to the fact that POSSE \cite{Gupta:2013} and SWUM \cite{HillSWUM:2010} use a limited tagset. Because SWUM/POSSE both rely on a more limited set, we use a manually curated mapping from the Penn Treebank set to our narrower tagset, provided in Table~\ref{tab:penntreebankmap}, which we now discuss. Many of these mappings take subcategories of various part-of-speech annotations and translate them to the broadest category. For example, proper noun $\longrightarrow$ noun and modal $\longrightarrow$ verb, where a proper noun is a more specific kind of noun, and a modal is just a more specific kind of verb.

Past tense verb (VBD), present participle verb (VBG), third-person verb (VBZ), and past participle verb (VBN) are all used as adjectives in some situations within our data set. For example, \textit{sortedIndicesBuf}, \textit{waitingList}, and \textit{adjustedGradient} where \textit{sorted} is a past tense verb (VBD), \textit{waiting} is a present participle (VBG), and \textit{adjusted} is a past participle verb (VBN). Next, when Stanford assigns List Item (LS) and Symbol (SYM) to words in our data set, it is typically mis-annotating nouns, so we map these to a noun. We discuss more about these issues in Section \ref{methodology}. Even with our mapping, there are a few annotations that one or more taggers do not support. POSSE does not support NPL, CJ, or PRE and groups P, PR, and DT under a single annotation; closed list. SWUM does not support NPL, VM, or CJ. Stanford does not support PRE. 

%Therefore, we process these twice for Stanford, which is the only tagger that detects these verb subcategories: once as verbs and once as adjectives to compare Stanford's accuracy under both configurations. VBZ (third-person verbs) are also sometimes used as adjectives, but based on our data this annotation is most frequently used for the words \textit{is} or \textit{equals}; typically in boolean identifiers or methods. Therefore, we treat VBZ as a verb since treating them as an adjective would generally result in lower accuracy. The same applies to VB and VBP; words which receive these annotations are typically verbs being used as verbs in our dataset.

\subsection{Noun, Verb, and Prepositional phrases} \label{phrasalconcepts}
There are a few linguistic concepts that come up when we analyze part-of-speech tagger output. Specifically, we will be dealing with noun phrases, verb phrases, and prepositional phrases. We define these terms and give an example. A Noun Phrase (NP) is a sequence of noun modifiers, such as noun-adjuncts and adjectives, followed by a noun, and optionally followed by other modifiers or prepositional phrases \cite{Manning:1999}. The noun in a noun phrase is typically referred to as a \textit{head-noun}; the entity which is being modified/described by the words to its left \cite{Deissenbock:2005} (or, for programmers, sometimes on either side of the noun-phrase) in the phrase. A Verb Phrase (VP) is a verb followed by an NP and optionally a Prepositional Phrase (PP). A PP is a preposition plus an NP. PPs can also be part of a larger VP or NP, as seen in Figure \ref{fig:example_phrasal_concepts}.

Figure~\ref{fig:example_phrasal_concepts} presents an example NP, VP, and VP with PP for three method name identifiers. The phrase structure nodes are NP, VP, and PP, while the other nodes (i.e., N, NM, V, P) are part-of-speech annotations. The leaf nodes are the individual words split from within the identifier. Each word in the identifier is assigned a part-of-speech, which can then be used to derive the identifier’s phrase structure. We cannot build a phrase structure without part-of-speech information. One important thing to note about these phrases is how the words in the phrases work together. For example, in noun phrases, noun modifiers (i.e., noun-adjuncts and adjectives) work to modify (i.e., specify) the concept represented by the head-noun that is part of the same phrase. In Figure~\ref{fig:example_phrasal_concepts}, \textit{contentBorder} is a noun phrase where \textit{content} modifies our understanding of the noun \textit{border}. It tells us that we are talking about the content border as opposed to another type of border; a \textit{window border}, for example. When we make it into a verb phrase by adding draw to get \textit{drawContentBorder}; we add an action (i.e., draw) that will be applied to the particular type of border (i.e., the content border) represented by the identifier.
\section{MOTIVATION \& RELATED WORK}
Part-of-speech tagging is an important analysis technique for understanding the meaning of words found in identifiers and the relationships between these words. Numerous prior works rely on some form of part-of-speech tagging to draw conclusions from identifier names \cite{Host:2009,Arnaoudova:2013,Peruma:2018:EIW:3242163.3242169, PERUMA2020110704, Perumascam,Newman2020, Butler:2009,butler:2011,butler:2015,Binkley:2011,Wu2020JSS,2021:ICPC:METHODS}. Thus, improved part-of-speech tagging technology can significantly improve the ability of researchers to understand the connection between identifier behavior and code behavior. Understanding this connection allows for stronger identifier name appraisal and suggestion as shown in prior research. For example, Host and Ostvold \cite{Host:2009} use part-of-speech tags to design a method of "debugging" (i.e., appraising) method names based partially on their part-of-speech sequence. Arnaoudova et al extends this idea by introducing linguistic anti-patterns \cite{Arnaoudova:2013}, and recent literature draws inspiration, in part, from these papers to create a way of checking method naming consistency using AI \cite{Nguyen:2020}. All based on an initial understanding that \textit{leveraged POS tagging}.

Arnaoudova et al also used part-of-speech tags to help categorize changes made to identifier names during rename operations \cite{Arnaoudova:2014}. Peruma et al relies on Arnaoudova's approach to understand the context around rename changes \cite{Peruma:2018:EIW:3242163.3242169, PERUMA2020110704, Perumascam} and identify patterns in the way names evolve. Prior work also relied on part-of-speech information to perform empirical analysis on identifiers in production code \cite{Newman2020, Butler:2009,butler:2011,butler:2015,Binkley:2011} to understand and taxonomize different naming structures and discuss how these varying structures are related to different types of program behavior. Part-of-speech information was also used to taxonomize test naming patterns and create test naming templates with a goal of constructing/improving tools to suggest and appraise test names \cite{Wu2020JSS, 2021:ICPC:METHODS}. All of this work relies on POS tagging or the concept of POS tags.

In essence, part-of-speech tagging is an important activity in the analysis of identifier names. Improved part-of-speech techniques will provide significant support for current and future research that seeks to understand identifier naming and its connection to source code behavior. It also helps these research tasks scale to larger and larger datasets. In the rest of this section we discuss the techniques above more specifically.

\label{sec:related}
\subsection{Part-of-speech taggers}
POSSE \cite{Gupta:2013} and SWUM \cite{HillSWUM:2010} are part-of-speech taggers created specifically to be run on software identifiers; they are trained to deal with the specialized context in which identifiers appear. Both POSSE and SWUM take advantage of static analysis to provide annotations. For example, they will look at the return type of a function to determine whether the word \textit{set} is a noun or a verb. Additionally, they are both aware of common naming structures in identifier names. For example, methods are more likely to contain a verb in certain positions within their name (e.g., at the beginning) \cite{Gupta:2013,HillSWUM:2010}. They leverage this information to help determine what POS to assign different words. Stanford \cite{Toutanova:StanfordTagger} is a popular POS tagger for general natural language (e.g., English) text. Olney et al. \cite{Olney2016} compared taggers for accuracy on 200+ identifiers, but only on Java method names. They found that SWUM and POSSE were the most accurate taggers for source code. Newman et al \cite{Newman2020} compared the same taggers but on a larger dataset (1,335 identifiers) and five identifier categories: function, class, attribute, parameter, and declaration statement. They found that SWUM was the most accurate overall, with an average accuracy around 59.4\% at the identifier level. Thus, there is significant room for improvement. We refer the reader to \cite{Newman2020} for a thorough evaluation and discussion of tagger accuracy on all types of identifiers. However, we provide some accuracy metrics for each tagger in our evaluation. 

Our dataset supports a larger tagset than POSSE and SWUM, but a narrower tagset (besides PRE) than Stanford, whose tagset is designed for standard human language sentence structure. Stanford supports all annotations except Preamble (PRE). POSSE does not support noun plurals (NPL), conjunctions (CJ), or preambles (PRE) and it groups prepositions (P), determiners (DT), and pronouns (PR) under the same annotation, which they call \textit{closed list}. We manually checked and determined that words annotated as \textit{closed list} were most frequently a preposition, thus we map these to preposition in POSSE's output. This may not always be correct, but there is no way to configure POSSE to provide the specific sub-annotation; it always annotates these as \textit{closed list}. Prior work \cite{Newman2020} shows that putting all of these under the same category may hinder analysis of the identifier and its code context, since determiners and prepositions provide useful hints about how an identifier is related to the code surrounding it. SWUM does not support noun plurals (NPL), adverbs (VM), or conjunctions (CJ). The fact that each tagger lacks support for one or more annotations does mean that their accuracy is lower on our dataset as a result.

Posit \cite{posit:2020} is an approach to tag mixed text; primarily motivated by the need to analyze emails and Stackoverflow text which may contain both code and standard natural language text. They use Stanford tagger \cite{Toutanova:StanfordTagger} as part of their approach, which is appropriate since they need to tag standard natural language sentences. This tagger is solving a somewhat different problem than ours, since it must differentiate code tokens from natural language tokens and then provide the correct categorization depending on which type the token falls under. Their technique specifically annotates code tokens with AST information rather than part-of-speech information. Our approaches are complementary; our tagger may be used by theirs to tag identifiers with both part-of-speech information to augment the AST information they provide.

\subsection{Part-of-speech-based Analysis of Identifiers}
Butler's work ~\cite{butler2010exploring} extends their previous work on Java class identifiers \cite{Butler:2009} to show that flawed method identifiers are also associated with low-quality code according to static analysis-based metrics. Caprile and Tonella \cite{tonella:1999} analyze the syntax and semantics of function identifiers. They create classes which can be used to understand the behavior of a function; grouping function identifiers by leveraging the words within them to understand some of the semantics of those identifiers. They also use the classes identified in this prior work to propose methods for restructuring program identifiers \cite{tonella:2000}. Fry and Shepherd \cite{Shepherd:2007,fry:2008} study verb-direct objects to link verbs to the natural-language-representation of the entity they act upon in order to assist in locating action-oriented concerns.

Butler \cite{butler:2011} studied class identifier names and lexical inheritance; analyzing the effect that interfaces or inheritance has on the name of a given class. For example, a class may inherit from a super class or implement a particular interface. Sometimes this class will incorporate words from the interface name or inherited class in its name. His study builds on work by Singer and Kirkham \cite{singer:2008}, who identified a grammar pattern for class names of (adjective)* (noun)+ and studies how class names correlate with micro patterns. Butler also studied Java field, argument, and variable naming structures \cite{butler:2015}. Among other results, they identify noun phrases as the most common pattern for field, argument, and variable names. Verb phrases are the second most common.

H{\o}st and {\O}stvold study method names as part of a line of work discussed in H{\o}st's disseration \cite{hostdissertation}. This line of work starts by analyzing a corpus of Java method implementations to establish the meanings of verbs in method names based on method behavior, which they measure using a set of attributes which they define \cite{HostLexicon}. They automatically create a lexicon of verbs that are commonly used by developers and a way to compare verbs in this lexicon by analyzing their program semantics. They build on this work in \cite{hostphrasebook} by using full method names which they refer to as phrases and augment their semantic model by considering a richer set of attributes. They extend this use of phrases by presenting an approach to debug method names \cite{Host:2009}. In this work, they designed automated naming rules using method signature elements. They use the phrase refinement from their prior paper, which takes a sequence of part-of-speech tags (i.e., phrases) and concretes them by substituting real words. (e.g., the phrase $<$verb$>$-$<$adjective$>$ might refine to is-empty). They connect these patterns to different method behaviors and use this to determine when a method's name and implementation do not match. They consider this a naming bug. Finally, in \cite{Hst2011CanonicalMN}, H{\o}st and {\O}stvold analyzed how ambiguous verbs in method names makes comprehension of Java programs more difficult. They proposed a way to detect when two or more verbs are synonymous.

Binkley et al. \cite{Binkley:2011} study grammar patterns for attribute names in classes. They come up with four rules for how to write attribute names: 1) Non-boolean field names should not contain a present tense verb, 2) field names should never only be a verb, 3) field names should never only be an adjective, and 4) boolean field names should contain a third person form of the verb “to be” or the auxiliary verb “should”.

Our work is complementary to the research above in that our improved tagger can be used to increase the quality of tools and techniques they present.
% \usepackage{booktabs}
\begin{table}[]
\centering
\caption{Total number per category of identifiers and unique grammar patterns across all systems}
\label{tab:systemstats}
\begin{tabular}{@{}lcc@{}}
\toprule
\multicolumn{1}{c}{Category} & \begin{tabular}[c]{@{}c@{}}Total Identifiers \\ Across All Systems\end{tabular} & \begin{tabular}[c]{@{}c@{}}\# of Unique Grammar\\ Patterns in Dataset\end{tabular} \\ \midrule
Decls & 920778 & 45 \\ \midrule
Classes & 37117 & 40 \\ \midrule
Functions & 428748 & 96 \\ \midrule
Parameters & 1197047 & 40 \\ \midrule
Attributes & 159562 & 53 \\ \midrule
\multicolumn{1}{c}{\textbf{Total}} & 2743252 & 277 \\ \bottomrule
\end{tabular}
\end{table}
\begin{table}[]
\centering
\caption{Distribution of Annotations in Training and Test Sets}
\label{tab:train_test_distribution}
\begin{tabular}{@{}lcc@{}}
\toprule
\multicolumn{1}{c}{Annotation} & Training Set & Unseen Test Set \\ \midrule
CJ & 11 & 1 \\ \midrule
D & 20 & 7 \\ \midrule
DT & 13 & 5 \\ \midrule
N & 1149 & 321 \\ \midrule
NM & 1520 & 414 \\ \midrule
NPL & 220 & 78 \\ \midrule
P & 91 & 32 \\ \midrule
PRE & 83 & 33 \\ \midrule
V & 330 & 81 \\ \midrule
VM & 12 & 3 \\ \midrule
\multicolumn{1}{c}{\textbf{Total}} & 3449 & 977 \\ \bottomrule
\end{tabular}
\end{table}

\section{Methodology}
\label{methodology}
In prior work \cite{Newman2020}, we constructed a dataset of 1,335 identifiers from 20 systems and manually annotated these identifiers with part-of-speech tags. For our evaluation of the ensemble tagger, we needed to create both a test set and a training set. We wanted the test set to contain identifiers from systems that were not present in the training set. Thus, we removed 5 systems and all corresponding identifiers from the original 20 system dataset and used these to create a test set. Since we wanted to maintain the same size as the original dataset, we collected additional identifiers from the remaining 15 systems so that the size of the dataset continued to be 1,335 identifiers. Thus, the training set used for our ensemble is derived from our original dataset \cite{Newman2020} but is not the exact same. The 15 system identifier set and the ensemble tagger are available through our webpage\footnote{https://scanl.org/} and on github\footnote{https://github.com/SCANL/ensemble\_tagger}.

Below, we explain our steps as if we collected and annotated the full 15 system dataset from scratch, as opposed to deriving it from prior work and expanding it, since these are all the steps used on all identifiers in our training and test sets. Explaining it this way simplifies the discussion. One thing to note is that we assess the quality of our model in multiple ways, meaning we have multiple test sets. We name these differently to ease the burden of reading. One set is called the "unseen test set" which is made up of identifiers from systems that were not in the training set. The other test set(s) are constructed using 5-fold cross validation, which splits the 15 system training set into smaller train-test sets. We call this the "5-fold test set". We will stick to this terminology for clarity throughout this section. 

\subsection{Training Set Construction}
We grouped identifiers into five categories: class names, function names, parameter names, attribute names (i.e., data members), and declaration-statement names. A declaration-statement name is a name belonging to a function-local or global variable. We use this terminology as it is consistent with srcML's terminology \cite{collard:2016} for these variables and we used srcML to collect identifiers.

This dataset includes 1,335 identifiers which break down into 3,449 words (Table \ref{tab:train_test_distribution}). The number was chosen by taking a sample from the total number of identifiers at 95 confidence level and 6 confidence interval from each of the five categories we support, meaning that we sampled 267 identifiers from each category (5*267=1,335). We collected our identifier set from 15 open-source systems. We chose these systems to vary in terms of size and programming language while also being mature and having their own development communities. We did this to make sure that the identifiers in these systems have been seen by multiple contributors and that the identifiers we collected are not biased toward a specific programming language. There are two reasons for choosing identifiers from multiple languages. 1) We want to know what patterns cross-cut between languages, such that most Java/C/C++ developers are familiar with and leverage these patterns. Focusing on just one language might mean the patterns we find are not common to developers outside of the chosen language. 2) Many systems are written in more than one language, and it is important to understand how well part-of-speech tagging technologies will work on these systems. Thus, running our study on systems written in different programming languages helps us study part-of-speech tagger results in an environment leveraging multiple programming languages.
\begin{table}[]
\centering
\caption{Systems used to create training (unbolded) and unseen test (\textbf{bolded}) sets.}
\label{tab:systemsizes}
\begin{tabular}{@{}lccc@{}}
\toprule
\multicolumn{1}{c}{\textbf{Name}} & \textbf{Size (kloc)} & \textbf{Age (years)} & \textbf{Language(s)} \\ \midrule
\textbf{junit4} & 30 & 19 & Java \\ \midrule
mockito & 46 & 9+ & Java \\ \midrule
\textbf{okhttp} & 54 & 6 & Java \\ \midrule
antlr4 & 92 & 27 & Java/C/C++/C\# \\ \midrule
\textbf{openFrameworks} & 130 & 14 & C/C++ \\ \midrule
jenkins & 156 & 8 & Java \\ \midrule
irrlicht & 250 & 13 & C/C++ \\ \midrule
kdevelop & 260 & 19 & C/C++ \\ \midrule
ogre & 370 & 14 & C/C++ \\ \midrule
quantlib & 370 & 19 & C/C++ \\ \midrule
\textbf{coreNLP} & 582 & 6 & Java \\ \midrule
swift & 601 & 5 & C++/C \\ \midrule
calligra & 660 & 19 & C/C++ \\ \midrule
gimp & 777 & 23 & C/C++ \\ \midrule
telegram & 912 & 6 & Java/C/C++ \\ \midrule
opencv & 1000 & 19 & C/C++ \\ \midrule
elasticsearch & 1300 & 9 & Java \\ \midrule
\textbf{bullet3} & 1300 & 10+ & C/C++/C\# \\ \midrule
blender & 1600 & 21 & C/C++ \\ \midrule
grpc & 1800 & 5 & C++/C/C\# \\ \bottomrule
\end{tabular}
\end{table}

We provide the list of systems and their characteristics in Table~\ref{tab:systemsizes}. The systems we picked were 615 KLOC on average with a median of 476 KLOC, a min of 30 KLOC, and a max of 1,800 KLOC. Further, most of these systems have been in active development for the past ten years or more and all of them for five years or more. The younger systems in our set are popular, modern programs. For example, Swift is a well-known programming language supported by Apple, Telegram is a popular messaging app, and Jenkins is a popular development automation server. Because we are trying to train an ensemble part-of-speech tagger to work well on as much code as possible, our goal is not necessarily to include only high-quality identifier names, but to include names that are closely representative of the average name for open-source systems. Additionally, we remove identifiers that appear in test files, in part because they sometimes have specialized naming conventions (e.g., include the word `test', `assert', `should', etc). This is supported by other research on test names \cite{Zhang2015ASE,Zhang2016ASE,Wu2020JSS}. We exclude test-related identifiers by ignoring annotated test files and directories; any directory, file, class, or function containing the word \textit{test}. While it is possible that identifiers in test code have similar grammar patterns to identifiers outside of test code, it is also possible that they do not. We did not want to risk introducing divergent grammar patterns. We think it would be appropriate to create a separate dataset of test identifier grammar patterns.

To collect the 1,335 identifiers, we scanned each of our 15 systems using srcML \cite{collard:2016} and collected both identifier names/types and the category that they fell into (e.g., class, function). Then, for each category, we randomly selected one identifier from each system using a round-robin algorithm (i.e., we picked a random identifier from system 1, then randomly selected an identifier from system 2, etc. until we hit 267). This ensured that we got either 17 or 18 identifiers from each system (267/15 = 17.8) per category and mitigates the threat of differing system size.

The dataset is balanced in terms of system (i.e., equal number of observations from each system) and in terms of code category (i.e., equal number of function names, parameter names, etc). However, the dataset is \textbf{not} balanced in terms of annotation. As shown in Table \ref{tab:train_test_distribution}, there are a different number of observations for each annotation we support in our tag set. We do not balance it for 3 reasons: 1) the unbalanced nature of the dataset mirrors reality more accurately; some annotations are much less common than others in English. 2) Balancing it would cause our data to be significantly different from the average distribution of identifier names \cite{Newman2020}. 3) Because no automated tagger is 100\% accurate, it would not be possible to automatically balance the dataset.

We did not expand abbreviations. The reason for this is that abbreviation expansion techniques are not widely available (e.g., cannot be easily integrated into different languages or frameworks, cannot be readily trained, are not fully or publicly implemented) and still not very accurate \cite{newmanabbrev}. Therefore, a realistic worst-case scenario for developers and researchers is that no abbreviation-expansion technique is available to use; their part-of-speech taggers must work in this worst-case scenario. We also tried not to split domain-term abbreviations (e.g., some splitters will make IPV4 into IPV 4; we corrected this back to IPV4). We did this because some taggers may recognize these domain terms. It is also the view of the authors that they should be recognized and appropriately tagged in their abbreviated (i.e., their most common) form.

As stated in Section \ref{annotations}, some verb forms are used as either adjectives or verbs. Stanford tagger is the only tagger that recognizes derivative verb forms such as past-tense or modal. Thus, it is the only one we need to configure. In prior work \cite{Newman2020} we measured how Stanford's accuracy is influenced in different contexts when we assume that verbs are being used as adjectives or verbs. In short, Stanford's accuracy increases when we assume that verb conjugations are adjectives in every context except function names. For function names, it is better to assume that its verb annotations are verbs. Thus, our training set reflects this reality. 

Finally, when we apply the Stanford tagger to function names, we append the letter \textit{I} to the beginning of the name. This is a known technique-- the Stanford+I technique, used to help Stanford tag identifiers that represent actions more accurately. It was used in previous studies applying part-of-speech tags to method identifier names \cite{Olney2016, Newman2020, Binkley:2011, abebe10} to increase Stanford's accuracy and confirmed to increase Stanford's accuracy on function names by Newman et al. \cite{Newman2020}.

% \begin{table}[]
% \centering
% \caption{Tukey HSD Test results for three independent taggers and top two ensemble taggers at the word-level. Bolded values are significant.}
% \label{tab:tukey_hsd_word}
% \textcolor{red}{
% \begin{tabular}{@{}l|cc@{}}
% \toprule
% \multicolumn{1}{c|}{Pair under test} & P-Value & Result \\ \midrule
% DTCP vs RFCP & 0.899995 & insignificant \\ \midrule
% DTCP vs SWUM & 0.050212 & insignificant \\ \midrule
% DTCP vs POSSE & \textbf{0.001005} & \textbf{p\textless{}0.01} \\ \midrule
% DTCP vs Stanford & \textbf{0.001005} & \textbf{p\textless{}0.01} \\ \midrule
% RFCP vs SWUM & \textbf{0.034793} & \textbf{p\textless{}0.05} \\ \midrule
% RFCP vs POSSE & \textbf{0.001005} & \textbf{p\textless{}0.01} \\ \midrule
% RFCP vs Stanford & \textbf{0.001005} & \textbf{p\textless{}0.01} \\ \midrule
% SWUM vs POSSE & \textbf{0.001005} & \textbf{p\textless{}0.01} \\ \midrule
% SWUM vs Stanford & \textbf{0.001005} & \textbf{p\textless{}0.01} \\ \midrule
% POSSE vs Stanford & 0.899995 & insignificant \\ \bottomrule
% \end{tabular}%
% }
% \end{table}

\subsection{Unseen Systems Test Set}
Our test set is made up of 384 identifiers that break down into 977 words (Table \ref{tab:train_test_distribution}) grouped by the same five categories used for the training set. It is constructed from identifiers contained in 5 systems that were removed from the original dataset \cite{Newman2020}, as explained at the beginning of Section \ref{methodology} and shown \textbf{bolded} in Table \ref{tab:systemsizes}. We based the number, 384, on a sample at 95 confidence level and 5 confidence interval. The population from which the sample was derived is the full set of identifiers across all categories. The size and breakdown of the full population, 2,743,252 identifiers, is found in Table \ref{tab:systemstats}. Given this sample size, we collected 76 or 77 (384/5 = 76.8) identifiers from each category. These were balanced for category as well as system (i.e., an equal number of identifiers from each system).  Like the training set, these identifiers were manually annotated with part-of-speech tags by three of the authors; each author taking a set to annotate on their own, and cross-validated by swapping sets to confirm (i.e., agree or disagree) that the manual annotation is correct. The authors came to full agreement on each identifier. 

There are multiple versions of this dataset. All of them have the same identifiers, but the annotations change based on which configuration was used to generate the set. The configurations come in pairs. One pair is \textit{normalized} or \textit{conjugated} and the other pair is \textit{augmented} or \textit{plain}. \textit{Normalized} datasets are those which convert all verb conjugations detected by standford to standard verbs (V in Table \ref{tab:posusedtable}). \textit{Conjugated} datasets are the opposite; they used all Stanford's verb conjugations. \textit{Augmented} datasets remove all annotations that have a frequency less than 25 and replaced them with an OTHER category. We use this to study whether rarely-seen part-of-speech tags have a negative effect on the overall quality of the tagger. \textit{Plain} datasets include all annotations shown in Table \ref{tab:posusedtable}.
\subsection{5-fold test set}
In addition to the unseen test set, we use k-fold cross validation to help us understand the generality of our ensemble tagging model. Typically, when using k-fold cross validation, prior researchers choose k as either 5 or 10. In this work, we choose 5 since it is most appropriate considering the distribution of annotations and the size of our dataset. The 5-fold test set is constructed from the training set of 1,335 identifiers. Effectively, the training set is split into five smaller sets. 30\% of these is chosen as a testing set and the other 70\% are used as training. Then, after training on four and testing on one, following typical 5-fold cross validation procedure, other sets are chosen as the test sets and the rest (now including some data that was just used as testing) are used to train. At each train-test step, we collect metrics about the effectiveness of our model as discussed in the next subsection.
% \begin{equation}
% \text{\footnotesize Precision}=\frac{\text{\footnotesize True Positive}}{\text{\footnotesize True Positive + False Positive}}
% \label{equation:precision}
% \end{equation}

% \begin{equation}
% \text{\footnotesize Recall}=\frac{\text{\footnotesize True Positive}}{\text{\footnotesize True Positive + False Negative}}
% \label{equation:recall}
% \end{equation}

% \begin{equation}
% \text{\footnotesize Accuracy}=\frac{\text{\footnotesize TrueNegative + TruePositive}}{\text{\footnotesize TrueNegative + FalsePositive + TruePositive + FalseNegative}
% }
% \label{equation:accuracy}
% \end{equation}

% \begin{equation}
% \text{\footnotesize F1 Score}=\text{\footnotesize 2} \cdot \frac{\text{\footnotesize Precision}\cdot \text{\footnotesize Recall}}{\text{\footnotesize Precision + Recall}}
% \label{equation:f1}
% \end{equation}
\subsection{Measuring Model Quality} \label{AccuracyDef}
We measure the quality of our ensemble using typical metrics for categorization problems. That is, we use Accuracy, Precision, Recall, and F1 Score. 
%We provide the equations we used to calculate these in equations \ref{equation:precision}, \ref{equation:recall}, \ref{equation:accuracy}, and \ref{equation:f1}. 
% \marcos{Do we need to provide P, R, F, and A equations? I would probably not include them in a paper submitted to an NLP journal/conference, but for this journal in particular, I don't know.}
% \marcos{About P, R, F, and A, it's clear from the the tables that they have been calculated for each class and then averaged. Should we make this information clear here? I might have missed it but I don't think we have stated that.}
In addition to the metrics above, we also report balanced accuracy in our 5-fold results, which is similar to accuracy except we calculate the average proportion of correct predictions for each individual annotation (i.e., N, NM, CJ, etc.) first and then divide by the number of annotations. Balanced accuracy helps when dealing with unbalanced datasets by giving more weight to annotations with lower frequency in the dataset.

\subsection{Choosing and training machine learning approaches}
For the evaluation of our ensemble, we chose to use Random Forest and Decision Tree as our primary machine learning approaches. %We did consider other approaches.
Initially, we considered Support Vector Classification, Logistic Regression, K-Nearest Neighbors, and Multinomial Naive Bayes. However, our preliminary analysis shows that Random Forest and Decision Tree outperforms the other classifiers in terms of our model quality metrics. Hence, we focus on evaluating the quality of our approach to using these two algorithms. To build our optimized model, we first split the dataset into a training and test set. The training set contains 70\% of the observations, while the remaining 30\% of the observations were part of the test set as validation data. To ensure that we are constructing an optimized model, we perform a hyperparameter optimization process. The purpose of this process is to evaluate a series of parameter values associated with the model to determine the set of values that result in the best performance of the model \cite{Albon201Machine}. Our hyperparameter tuning process involved an exhaustive grid search \cite{dangeti2017statistics} and 5-fold cross-validation on the training dataset. Grid search utilizes a brute force technique to evaluate all combinations of hyperparameters to obtain the best performance. Provided in Table \ref{Table:hyperparameter}, are the optimal hyperparameter values for the classification algorithms in our study.
%We used 5-fold validation to determine which of these algorithms generally performed the best in terms of the quality metrics discussed above in addition to speed. Decision Tree and Random Forest both outperformed all other approaches, so instead of continuing to include these other algorithms, we focused on evaluating the quality of our approach when it is using one of these two algorithms.

\begin{table}
\centering
\caption{Optimal parameter values for the classification algorithms.}
\label{Table:hyperparameter}
\begin{tabular}{@{}lll@{}}
\toprule
\multicolumn{1}{l}{\textbf{Algorithm}} & \multicolumn{1}{l}{\textbf{Parameter}} & \multicolumn{1}{l}{\textbf{Value}} \\ \midrule
\multirow{4}{*}{Random Forest} & max\_depth & 83 \\
 & n\_estimators & 250 \\
 & criterion & gini \\
 & bootstrap & True \\ \midrule
\multirow{2}{*}{Decision Tree} & criterion & entropy \\
 & max\_depth & 9 \\ 
\bottomrule
\end{tabular}
\end{table}

\subsection{Dataset preparation and Features}\label{prepandfeatures}
To prepare the datasets for annotation by the ensemble, we run SWUM \cite{HillSWUM:2010}, POSSE \cite{Gupta:2013}, and Stanford \cite{Toutanova:StanfordTagger} on each identifier to obtain their individual annotations. We provide any information required by the three taggers above (e.g., SWUM requires function signatures). Since the dataset is pre-split from prior work \cite{Newman2020}, we do not have to worry about splitting. In addition, as stated prior, the correct annotation is already provided. After we have run these three taggers on the data, we vectorize the data by splitting all identifiers into their individual words along with the part-of-speech provided to them by SWUM, POSSE, Stanford, and the human annotators. In addition, we collect several data characteristics to serve as features to help the ensemble correctly annotate the data. We explain these characteristics now.

Machine learning algorithms use characteristics of the data they are trained on to learn the nuances of that data such that they are able to use these characteristics to categorize unseen data. These characteristics are typically called Features. Our ensemble uses several features to annotate (i.e., categorize) identifiers with part of speech tags. The features that we considered for our model are based on empirical results from a prior study we performed on the grammar patterns latent in source code identifiers \cite{Newman2020}. To summarize, we found that certain annotations are heavily correlated with 1) words that appear in certain positions. For example, nouns appear at the end of an identifier, noun-adjuncts appear at the beginning or middle; 2) with the type of an identifier. For example, verbs are more common in boolean-type identifiers; and 3) certain contexts. For example, verb phrases are more common in function names. We also noticed that certain taggers are better at recognizing certain annotations than others. For example, SWUM is great at recognizing noun-adjuncts, Stanford is great at recognizing conjunctions and prepositions. Therefore, we chose features that will help our ensemble take advantage of these patterns. Most of these features are also very easy to obtain using static analysis, making them very accessible in different environments and, thus, helping guarantee ease of integrating our approach into another applications. The features that we considered for our model are as follows:

\begin{enumerate}
    \item \textbf{Word} - The word itself.
    \item \textbf{Data Type} - the type (or return type) of an identifier (or function identifier).
    \item \textbf{SWUM annotation} - The annotation that the SWUM POS tagger applied to a given word.
    \item \textbf{POSSE annotation} - The annotation that the POSSE POS tagger applied to a given word.
    \item \textbf{Stanford annotation} - The annotation that the Stanford POS tagger applied to a given word.
    \item \textbf{Position} - The position of a given word within its original identifier. For example, given an identifier: \textit{GetXMLReaderHandler}, \textit{Get} is in position 1, \textit{XML} is in position 2, \textit{Reader} is in position 3 and \textit{Handler} is in position 4.
    \item \textbf{Identifier size} - The length, in words, of the identifier of which the word was originally part. 
    \item \textbf{Normalized position} - We normalized the position metric described above such that the first word in the identifier is in position 1, all middle words are in position 2, and the last word is in position 3. For example, given an identifier: \textit{GetXMLReaderHandler}, \textit{Get} is in position 1, \textit{XML} is in position 2, \textit{Reader} is in position 2 and \textit{Handler} is in position 3. The reason for this feature is to mitigate the sometimes-negative effect of very long identifiers.
    \item \textbf{Context} - The dataset contains five categories of identifier name: function, parameter, attribute, declaration, and class. We provide the category to which the given identifier belongs as one of the features to allow the ensemble to learn patterns that are more pervasive for certain identifier types versus others. For example, function identifiers contain verbs at a higher rate \cite{Newman2020,Gupta:2013, HillSWUM:2010, Host:2009} than other types of identifiers.
\end{enumerate}
%\decker{Something to try including latter, inverse position.  0 is right most, -1 is second right-most, etc}
We tested all of these features using 5-fold cross validation and the metrics Described in Section \ref{AccuracyDef}. Specifically, we were trying to determine what set of features maximized F1, Accuracy, and Balanced accuracy. To do this, we used 2 techniques: \textbf{Drop-column feature importance} and \textbf{permutation importance}. Drop-column feature importance can by calculated by creating a power set (i.e., all subsets) of the full set of features and then retraining your model with each subset. In this way, we can consider every possible subset of features for our feature set and determine which subset gives us the best performance with respect to F1, Accuracy, and Balanced Accuracy. While performing Drop-column feature importance, we also performed permutation importance. Permutation importance is done after a model has been fitted. It is defined as the decrease in a model score (i.e., our metrics) when a single feature's value is randomly shuffled. In essence, it measures how our metrics change when a feature is shuffled. Thus, for each subset of features in our ensemble, we also measure permutation importance.

Since there are a lot of subsets (i.e., power set of our 9 features is $2^9$ = 512), we only present data about the best feature set: \textbf{SWUM}, \textbf{POSSE}, and \textbf{Stanford} annotations, \textbf{Normalized position}, and \textbf{Context}. In addition, we present the permutation importances for these features in Tables \ref{tab:DTfeatureImportances} and \ref{tab:RFfeatureImportances}. These tables correspond to permutation importances for our best Random-Forest-based ensemble and our best Decision-Tree-based ensemble. In each table, you can see how each of the best features influenced F1, Balanced Accuracy, and Accuracy. Since we used 5-fold validation, there are 5 values in each row followed by an average of those values. \textit{A higher number means a feature is more important.} In general, out of our three taggers, SWUM had the highest influence on F1 and Accuracy, while Stanford had the highest influence on Balanced Accuracy. Of the non-tagger features (i.e., Normalized Position and Context), Normalized Position had the highest influence on all three metrics. We discuss more about why these features were most important in the Evaluation Section (Section \ref{evaluation}).

Finally, the features that were removed: Word, data type, position, and identifier size degraded our model performance. Our prior work \cite{Newman2020} gives us some insight as to why this might be. Starting with \textit{position} and \textit{max position}, we found that verb and noun phrases tended to begin with a particular annotation; a verb or noun modifier respectively. They also ended with a specific annotation: a Noun (i.e., a head-noun). Between the starting verb/noun modifier and the ending head-noun are a sequence of Noun Modifiers. Notice that this correlates to a \textit{beginning}, \textit{middle}, \textit{end} structure where the first word has a specific tag, all the middle words have the same tag, and the final word has a specific tag. Position and Max position confuse the ensemble because identifiers have varying lengths. Normalized position categorizes position as beginning, middle, or end. Thus, it improves the performance of the ensemble whereas position and max position hurt the performance. Word and data type can help the model recognize certain words and their correlation to different tags. However, on unseen data this may cause the ensemble to become confused because it sees words that it has not seen before. Thus, the ensemble will tag common words more accurately with these features turned on but uncommon/unseen words less accurately.

\begin{table}[]
\centering
\caption{Decision Tree feature importances for best features}
\label{tab:DTfeatureImportances}
\begin{tabular}{@{}lcccccc@{}}
\toprule
\textbf{Feature Set} & \multicolumn{5}{c}{F1 Weighted Importances} & Average \\ \midrule
SWUM & 0.26 & 0.26 & 0.25 & 0.26 & \multicolumn{1}{c|}{0.26} & 0.26 \\
POSSE & 0.15 & 0.14 & 0.15 & 0.15 & \multicolumn{1}{c|}{0.15} & 0.15 \\
Stanford & 0.23 & 0.23 & 0.23 & 0.23 & \multicolumn{1}{c|}{0.23} & 0.23 \\
Context & 0.02 & 0.02 & 0.02 & 0.02 & \multicolumn{1}{c|}{0.02} & 0.02 \\
Nomalized Position & 0.13 & 0.13 & 0.12 & 0.13 & \multicolumn{1}{c|}{0.13} & 0.13 \\ \midrule
\textbf{Feature Set} & \multicolumn{5}{c}{Balanced Accuracy Importances} & Average \\ \midrule
SWUM & 0.29 & 0.27 & 0.25 & 0.28 & \multicolumn{1}{c|}{0.27} & 0.27 \\
POSSE & 0.21 & 0.21 & 0.21 & 0.24 & \multicolumn{1}{c|}{0.21} & 0.22 \\
Stanford & 0.51 & 0.51 & 0.49 & 0.53 & \multicolumn{1}{c|}{0.51} & 0.51 \\
Context & 0.05 & 0.07 & 0.05 & 0.07 & \multicolumn{1}{c|}{0.07} & 0.06 \\
Nomalized Position & 0.10 & 0.10 & 0.08 & 0.10 & \multicolumn{1}{c|}{0.08} & 0.09 \\ \midrule
\textbf{Feature Set} & \multicolumn{5}{c}{Accuracy Importances} & Average \\ \midrule
SWUM & 0.26 & 0.26 & 0.26 & 0.26 & \multicolumn{1}{c|}{0.26} & 0.26 \\
POSSE & 0.14 & 0.13 & 0.13 & 0.13 & \multicolumn{1}{c|}{0.13} & 0.13 \\
Stanford & 0.22 & 0.21 & 0.20 & 0.20 & \multicolumn{1}{c|}{0.21} & 0.21 \\
Context & 0.02 & 0.02 & 0.02 & 0.02 & \multicolumn{1}{c|}{0.02} & 0.02 \\
Nomalized Position & 0.13 & 0.13 & 0.14 & 0.13 & 0.14 & 0.13 \\ \bottomrule
\end{tabular}
\end{table}

\begin{table}[]
\centering
\caption{Random Forest feature importances for best features}
\label{tab:RFfeatureImportances}

\begin{tabular}{@{}lcccccc@{}}
\toprule
\textbf{Feature Set} & \multicolumn{5}{c}{\textbf{F1 Weighted Importances}} & \textbf{Average} \\ \midrule
SWUM & 0.22 & 0.22 & 0.21 & 0.21 & \multicolumn{1}{c|}{0.22} & 0.21 \\
POSSE & 0.14 & 0.15 & 0.15 & 0.14 & \multicolumn{1}{c|}{0.15} & 0.14 \\
Stanford & 0.21 & 0.21 & 0.22 & 0.21 & \multicolumn{1}{c|}{0.21} & 0.21 \\
Context & 0.03 & 0.03 & 0.03 & 0.03 & \multicolumn{1}{c|}{0.03} & 0.03 \\
Nomalized Position & 0.17 & 0.16 & 0.15 & 0.16 & \multicolumn{1}{c|}{0.15} & 0.16 \\ \midrule
\textbf{Feature Set} & \multicolumn{5}{c}{\textbf{Balanced Accuracy Importances}} & \textbf{Average} \\ \midrule
SWUM & 0.22 & 0.22 & 0.21 & 0.26 & \multicolumn{1}{c|}{0.25} & 0.23 \\
POSSE & 0.23 & 0.23 & 0.20 & 0.23 & \multicolumn{1}{c|}{0.21} & 0.22 \\
Stanford & 0.47 & 0.47 & 0.50 & 0.50 & \multicolumn{1}{c|}{0.49} & 0.48 \\
Context & 0.05 & 0.04 & 0.07 & 0.06 & \multicolumn{1}{c|}{0.06} & 0.06 \\
Nomalized Position & 0.15 & 0.12 & 0.15 & 0.19 & \multicolumn{1}{c|}{0.21} & 0.17 \\ \midrule
\textbf{Feature Set} & \multicolumn{5}{c}{\textbf{Accuracy Importances}} & \textbf{Average} \\ \midrule
SWUM & 0.23 & 0.22 & 0.21 & 0.22 & \multicolumn{1}{c|}{0.22} & 0.22 \\
POSSE & 0.14 & 0.13 & 0.13 & 0.13 & \multicolumn{1}{c|}{0.13} & 0.13 \\
Stanford & 0.19 & 0.18 & 0.19 & 0.19 & \multicolumn{1}{c|}{0.19} & 0.19 \\
Context & 0.03 & 0.03 & 0.03 & 0.02 & \multicolumn{1}{c|}{0.02} & 0.03 \\
Nomalized Position & 0.16 & 0.16 & 0.15 & 0.16 & 0.16 & 0.16 \\ \bottomrule
\end{tabular}
\end{table}
\section{Evaluation Setup}\label{setup}
The dataset described in Section \ref{methodology} has several configurations that we use during evaluation. These configurations are as follows:
\begin{enumerate}
    \item The type of machine learning approach used; either Decision Tree or Random forest. They have the codes DT and RF respectively.
    \item The version of the dataset being used; either the plain dataset or the augmented dataset. These have the codes P and A respectively.
    \item Whether or not the Stanford data within the dataset is using verb conjugations or is normalized. These have the codes C and N respectively.
\end{enumerate}
%\decker{Makes sense to be.  Might be slightly better (not necessary) to put a - between ML and rest.  RF-CP}
To determine which configuration you are looking at when reading our results, look at the code present in each table. For example, if some data in a table has the code RFCP, then it used \textbf{random forest} with \textbf{conjugated} stanford identifiers and the \textbf{plain} dataset.
\begin{table*}[]
\centering
\caption{Five-fold validation results for each machine learning approach and configuration using the augmented dataset}
\label{tab:five_fold_results_other}
\begin{tabular}{@{}lcccccccccccc@{}}
\toprule
 & \multicolumn{5}{c}{\textbf{DTNA}} & \multicolumn{1}{l}{\textbf{Average}} & \multicolumn{5}{c}{\textbf{RFNA}} & \multicolumn{1}{l}{\textbf{Average}} \\ \midrule
\multicolumn{1}{l|}{Accuracy} & 0.80 & 0.84 & 0.79 & 0.84 & \multicolumn{1}{c|}{0.82} & \multicolumn{1}{c|}{\textbf{0.82}} & 0.81 & 0.83 & 0.79 & 0.84 & \multicolumn{1}{c|}{0.84} & \multicolumn{1}{c}{\textbf{0.82}} \\ \midrule
\multicolumn{1}{l|}{Balanced Accuracy} & 0.54 & 0.68 & 0.65 & 0.66 & \multicolumn{1}{c|}{0.59} & \multicolumn{1}{c|}{0.62} & 0.57 & 0.66 & 0.71 & 0.65 & \multicolumn{1}{c|}{0.75} & \multicolumn{1}{c}{\textbf{0.67}} \\ \midrule
\multicolumn{1}{l|}{Weighted F1} & 0.79 & 0.83 & 0.79 & 0.83 & \multicolumn{1}{c|}{0.81} & \multicolumn{1}{c|}{0.81} & 0.80 & 0.82 & 0.79 & 0.84 & \multicolumn{1}{c|}{0.84} & \multicolumn{1}{c}{\textbf{0.82}} \\ \midrule
\multicolumn{1}{l|}{Weighted Precision} & 0.80 & 0.82 & 0.80 & 0.83 & \multicolumn{1}{c|}{0.82} & \multicolumn{1}{c|}{0.81} & 0.81 & 0.82 & 0.80 & 0.84 & \multicolumn{1}{c|}{0.84} & \multicolumn{1}{c}{\textbf{0.82}} \\ \midrule
\multicolumn{1}{l|}{Weighted Recall} & 0.80 & 0.84 & 0.79 & 0.84 & \multicolumn{1}{c|}{0.82} & \multicolumn{1}{c|}{\textbf{0.82}} & 0.81 & 0.83 & 0.79 & 0.84 & \multicolumn{1}{c|}{0.84} & \multicolumn{1}{c}{\textbf{0.82}} \\ \midrule
 & \multicolumn{5}{c}{\textbf{DTCA}} & \multicolumn{1}{l}{\textbf{Average}} & \multicolumn{5}{c}{\textbf{RFCA}} & \multicolumn{1}{l}{\textbf{Average}} \\ \midrule
\multicolumn{1}{l|}{Accuracy} & 0.82 & 0.85 & 0.81 & 0.84 & \multicolumn{1}{c|}{0.84} & \multicolumn{1}{c|}{0.83} & 0.84 & 0.86 & 0.81 & 0.85 & \multicolumn{1}{c|}{0.86} & \multicolumn{1}{c}{\textbf{0.84}} \\ \midrule
\multicolumn{1}{l|}{Balanced Accuracy} & 0.51 & 0.62 & 0.55 & 0.53 & \multicolumn{1}{c|}{0.58} & \multicolumn{1}{c|}{0.56} & 0.52 & 0.69 & 0.62 & 0.48 & \multicolumn{1}{c|}{0.65} & \multicolumn{1}{c}{\textbf{0.59}} \\ \midrule
\multicolumn{1}{l|}{Weighted F1} & 0.82 & 0.84 & 0.80 & 0.82 & \multicolumn{1}{c|}{0.82} & \multicolumn{1}{c|}{0.82} & 0.83 & 0.86 & 0.80 & 0.84 & \multicolumn{1}{c|}{0.86} & \multicolumn{1}{c}{\textbf{0.84}} \\ \midrule
\multicolumn{1}{l|}{Weighted Precision} & 0.83 & 0.84 & 0.81 & 0.82 & \multicolumn{1}{c|}{0.82} & \multicolumn{1}{c|}{0.82} & 0.83 & 0.87 & 0.80 & 0.85 & \multicolumn{1}{c|}{0.86} & \multicolumn{1}{c}{\textbf{0.84}} \\ \midrule
\multicolumn{1}{l|}{Weighted Recall} & 0.82 & 0.85 & 0.81 & 0.84 & \multicolumn{1}{c|}{0.84} & \multicolumn{1}{c|}{0.83} & 0.84 & 0.86 & 0.81 & 0.85 & \multicolumn{1}{c|}{0.86} & \multicolumn{1}{c}{\textbf{0.84}} \\ \bottomrule
\end{tabular}%
\end{table*}
\begin{table*}[]
\centering
\caption{Five-fold validation results for each machine learning approach and configuration using the plain dataset}
\label{tab:five_fold_results_plain}
\begin{tabular}{@{}lcccccccccccc@{}}
\toprule
 & \multicolumn{5}{c}{\textbf{DTNP}} & \multicolumn{1}{l}{\textbf{Average}} & \multicolumn{5}{c}{\textbf{RFNP}} & \multicolumn{1}{l}{\textbf{Average}} \\ \midrule
\multicolumn{1}{l|}{Accuracy} & 0.81 & 0.81 & 0.86 & 0.82 & \multicolumn{1}{c|}{0.82} & \multicolumn{1}{c|}{0.82} & 0.84 & 0.82 & 0.87 & 0.82 & \multicolumn{1}{c|}{0.85} & \multicolumn{1}{c}{\textbf{0.84}} \\ \midrule
\multicolumn{1}{l|}{Balanced Accuracy} & 0.54 & 0.59 & 0.71 & 0.75 & \multicolumn{1}{c|}{0.53} & \multicolumn{1}{c|}{0.62} & 0.61 & 0.60 & 0.76 & 0.74 & \multicolumn{1}{c|}{0.60} & \multicolumn{1}{c}{\textbf{0.66}} \\ \midrule
\multicolumn{1}{l|}{Weighted F1} & 0.79 & 0.80 & 0.86 & 0.81 & \multicolumn{1}{c|}{0.80} & \multicolumn{1}{c|}{0.81} & 0.82 & 0.82 & 0.87 & 0.81 & \multicolumn{1}{c|}{0.84} & \multicolumn{1}{c}{\textbf{0.83}} \\ \midrule
\multicolumn{1}{l|}{Weighted Precision} & 0.80 & 0.80 & 0.87 & 0.82 & \multicolumn{1}{c|}{0.79} & \multicolumn{1}{c|}{0.82} & 0.82 & 0.82 & 0.87 & 0.81 & \multicolumn{1}{c|}{0.85} & \multicolumn{1}{c}{\textbf{0.83}} \\ \midrule
\multicolumn{1}{l|}{Weighted Recall} & 0.81 & 0.81 & 0.86 & 0.82 & \multicolumn{1}{c|}{0.82} & \multicolumn{1}{c|}{0.82} & 0.84 & 0.82 & 0.87 & 0.82 & \multicolumn{1}{c|}{0.85} & \multicolumn{1}{c}{\textbf{0.84}} \\ \midrule
 & \multicolumn{5}{c}{\textbf{DTCP}} & \multicolumn{1}{l}{\textbf{Average}} & \multicolumn{5}{c}{\textbf{RFCP}} & \multicolumn{1}{l}{\textbf{Average}} \\ \midrule
\multicolumn{1}{l|}{Accuracy} & 0.85 & 0.82 & 0.83 & 0.84 & \multicolumn{1}{c|}{0.85} & \multicolumn{1}{c|}{\textbf{0.84}} & 0.86 & 0.83 & 0.83 & 0.85 & \multicolumn{1}{c|}{0.85} & \multicolumn{1}{c}{\textbf{0.84}} \\ \midrule
\multicolumn{1}{l|}{Balanced Accuracy} & 0.58 & 0.67 & 0.73 & 0.60 & \multicolumn{1}{c|}{0.63} & \multicolumn{1}{c|}{\textbf{0.64}} & 0.51 & 0.59 & 0.54 & 0.50 & \multicolumn{1}{c|}{0.63} & \multicolumn{1}{c}{0.55} \\ \midrule
\multicolumn{1}{l|}{Weighted F1} & 0.85 & 0.82 & 0.83 & 0.84 & \multicolumn{1}{c|}{0.84} & \multicolumn{1}{c|}{0.83} & 0.86 & 0.82 & 0.82 & 0.84 & \multicolumn{1}{c|}{0.84} & \multicolumn{1}{c}{\textbf{0.84}} \\ \midrule
\multicolumn{1}{l|}{Weighted Precision} & 0.85 & 0.81 & 0.83 & 0.83 & \multicolumn{1}{c|}{0.83} & \multicolumn{1}{c|}{\textbf{0.83}} & 0.86 & 0.82 & 0.82 & 0.83 & \multicolumn{1}{c|}{0.83} & \multicolumn{1}{c}{\textbf{0.83}} \\ \midrule
\multicolumn{1}{l|}{Weighted Recall} & 0.85 & 0.82 & 0.83 & 0.84 & \multicolumn{1}{c|}{0.85} & \multicolumn{1}{c|}{\textbf{0.84}} & 0.86 & 0.83 & 0.83 & 0.85 & \multicolumn{1}{c|}{0.85} & \multicolumn{1}{c}{\textbf{0.84}} \\ \bottomrule
\end{tabular}%
\end{table*}
\begin{table*}[]
\centering
\caption{Per-annotation and Overall Accuracy of Ensemble Tagger on Augmented Dataset}
\label{tab:othertable}
\resizebox{\textwidth}{!}{%
\begin{tabular}{@{}llcccccccccccccccc@{}}
\toprule
 &  & \multicolumn{4}{c}{\textbf{DTNA}} & \multicolumn{4}{c}{\textbf{RFNA}} & \multicolumn{4}{c}{\textbf{DTCA}} & \multicolumn{4}{c}{\textbf{RFCA}} \\ \midrule
\multicolumn{1}{l|}{Annotation} & \multicolumn{1}{l|}{Total} & \multicolumn{1}{l}{Precision} & \multicolumn{1}{l}{Recall} & \multicolumn{1}{l}{F1} & \multicolumn{1}{l|}{Total} & \multicolumn{1}{l}{Precision} & \multicolumn{1}{l}{Recall} & \multicolumn{1}{l}{F1} & \multicolumn{1}{l|}{Total} & \multicolumn{1}{l}{Precision} & \multicolumn{1}{l}{Recall} & \multicolumn{1}{l}{F1} & \multicolumn{1}{l|}{Total} & \multicolumn{1}{l}{Precision} & \multicolumn{1}{l}{Recall} & \multicolumn{1}{l}{F1} & \multicolumn{1}{l}{Total} \\ \midrule
\multicolumn{1}{l|}{N} & \multicolumn{1}{l|}{322} & 0.87 & 0.89 & \textbf{0.88} & \multicolumn{1}{c|}{329} & 0.86 & 0.89 & \textbf{0.88} & \multicolumn{1}{c|}{331} & 0.84 & 0.89 & 0.87 & \multicolumn{1}{c|}{340} & 0.87 & 0.90 & \textbf{0.88} & 332 \\ \midrule
\multicolumn{1}{l|}{NM} & \multicolumn{1}{l|}{415} & 0.85 & 0.92 & \textbf{0.89} & \multicolumn{1}{c|}{448} & 0.85 & 0.92 & \textbf{0.89} & \multicolumn{1}{c|}{446} & 0.87 & 0.91 & \textbf{0.89} & \multicolumn{1}{c|}{435} & 0.86 & 0.92 & \textbf{0.89} & 447 \\ \midrule
\multicolumn{1}{l|}{NPL} & \multicolumn{1}{l|}{78} & 0.90 & 0.68 & 0.77 & \multicolumn{1}{c|}{59} & 0.93 & 0.67 & 0.78 & \multicolumn{1}{c|}{56} & 0.89 & 0.72 & 0.79 & \multicolumn{1}{c|}{63} & 0.96 & 0.69 & \textbf{0.81} & 56 \\ \midrule
\multicolumn{1}{l|}{OTHER} & \multicolumn{1}{l|}{16} & 0.69 & 0.56 & 0.62 & \multicolumn{1}{c|}{13} & 0.73 & 0.69 & \textbf{0.71} & \multicolumn{1}{c|}{15} & 0.56 & 0.31 & 0.40 & \multicolumn{1}{c|}{9} & 0.50 & 0.25 & 0.33 & 8 \\ \midrule
\multicolumn{1}{l|}{P} & \multicolumn{1}{l|}{32} & 0.70 & 0.81 & 0.75 & \multicolumn{1}{c|}{37} & 0.71 & 0.84 & 0.77 & \multicolumn{1}{c|}{38} & 0.68 & 0.78 & 0.72 & \multicolumn{1}{c|}{37} & 0.72 & 0.88 & \textbf{0.79} & 39 \\ \midrule
\multicolumn{1}{l|}{PRE} & \multicolumn{1}{l|}{33} & 0.64 & 0.21 & 0.32 & \multicolumn{1}{c|}{11} & 0.64 & 0.21 & 0.32 & \multicolumn{1}{c|}{11} & 0.78 & 0.21 & 0.33 & \multicolumn{1}{c|}{9} & 0.77 & 0.30 & \textbf{0.43} & 13 \\ \midrule
\multicolumn{1}{l|}{V} & \multicolumn{1}{l|}{81} & 0.79 & 0.78 & 0.78 & \multicolumn{1}{c|}{80} & 0.79 & 0.78 & 0.78 & \multicolumn{1}{c|}{80} & 0.79 & 0.81 & 0.80 & \multicolumn{1}{c|}{84} & 0.80 & 0.81 & \textbf{0.81} & 82 \\ \midrule
\multicolumn{1}{r}{\textbf{Accuracy}} & \multicolumn{1}{r}{} & \multicolumn{4}{c}{\textbf{0.84}} & \multicolumn{4}{c}{\textbf{0.85}} & \multicolumn{4}{c}{\textbf{0.84}} & \multicolumn{4}{c}{\textbf{0.85}} \\ \bottomrule
\end{tabular}
}
\end{table*}

\begin{table*}[]
\centering
\caption{Per-annotation and Overall Accuracy of Ensemble Tagger on Plain Dataset}
\label{tab:plaintable}
\resizebox{\textwidth}{!}{%
\begin{tabular}{@{}lccccccccccccccccc@{}}
\toprule
 & \multicolumn{1}{l}{} & \multicolumn{4}{c}{\textbf{DTNP}} & \multicolumn{4}{c}{\textbf{RFNP}} & \multicolumn{4}{c}{\textbf{DTCP}} & \multicolumn{4}{c}{\textbf{RFCP}} \\ \midrule
\multicolumn{1}{l|}{Annotation} & \multicolumn{1}{l|}{Total} & Precision & Recall & F1 & \multicolumn{1}{c|}{Total} & Precision & Recall & F1 & \multicolumn{1}{c|}{Total} & Precision & Recall & F1 & \multicolumn{1}{c|}{Total} & Precision & Recall & F1 & Total \\ \midrule
\multicolumn{1}{l|}{CJ} & \multicolumn{1}{c|}{1} & 0.50 & 1.00 & 0.67 & \multicolumn{1}{c|}{2} & 1.00 & 1.00 & \textbf{1.00} & \multicolumn{1}{c|}{1} & 1.00 & 1.00 & \textbf{1.00} & \multicolumn{1}{c|}{1} & 1.00 & 1.00 & \textbf{1.00} & 1 \\ \midrule
\multicolumn{1}{l|}{D} & \multicolumn{1}{c|}{7} & 0.88 & 1.00 & \textbf{0.93} & \multicolumn{1}{c|}{8} & 0.88 & 1.00 & \textbf{0.93} & \multicolumn{1}{c|}{8} & 0.88 & 1.00 & \textbf{0.93} & \multicolumn{1}{c|}{8} & 0.88 & 1.00 & \textbf{0.93} & 8 \\ \midrule
\multicolumn{1}{l|}{DT} & \multicolumn{1}{c|}{5} & 1.00 & 0.20 & 0.33 & \multicolumn{1}{c|}{1} & 1.00 & 0.40 & 0.57 & \multicolumn{1}{c|}{2} & 1.00 & 0.60 & \textbf{0.75} & \multicolumn{1}{c|}{3} & 1.00 & 0.60 & \textbf{0.75} & 3 \\ \midrule
\multicolumn{1}{l|}{N} & \multicolumn{1}{c|}{322} & 0.86 & 0.90 & 0.88 & \multicolumn{1}{c|}{338} & 0.87 & 0.89 & 0.88 & \multicolumn{1}{c|}{328} & 0.85 & 0.90 & 0.88 & \multicolumn{1}{c|}{340} & 0.88 & 0.89 & \textbf{0.89} & 327 \\ \midrule
\multicolumn{1}{l|}{NM} & \multicolumn{1}{c|}{415} & 0.85 & 0.93 & \textbf{0.89} & \multicolumn{1}{c|}{452} & 0.85 & 0.93 & \textbf{0.89} & \multicolumn{1}{c|}{453} & 0.88 & 0.91 & \textbf{0.89} & \multicolumn{1}{c|}{430} & 0.87 & 0.92 & \textbf{0.89} & 438 \\ \midrule
\multicolumn{1}{l|}{NPL} & \multicolumn{1}{c|}{78} & 0.98 & 0.65 & 0.78 & \multicolumn{1}{c|}{52} & 0.93 & 0.67 & 0.78 & \multicolumn{1}{c|}{56} & 0.90 & 0.72 & \textbf{0.80} & \multicolumn{1}{c|}{62} & 0.93 & 0.69 & 0.79 & 58 \\ \midrule
\multicolumn{1}{l|}{P} & \multicolumn{1}{c|}{32} & 0.65 & 0.81 & 0.72 & \multicolumn{1}{c|}{40} & 0.72 & 0.88 & \textbf{0.79} & \multicolumn{1}{c|}{39} & 0.68 & 0.84 & 0.75 & \multicolumn{1}{c|}{40} & 0.70 & 0.88 & 0.78 & 40 \\ \midrule
\multicolumn{1}{l|}{PRE} & \multicolumn{1}{c|}{33} & 0.70 & 0.21 & 0.33 & \multicolumn{1}{c|}{10} & 0.64 & 0.21 & 0.32 & \multicolumn{1}{c|}{11} & 0.78 & 0.21 & 0.33 & \multicolumn{1}{c|}{9} & 0.77 & 0.30 & \textbf{0.43} & 13 \\ \midrule
\multicolumn{1}{l|}{V} & \multicolumn{1}{c|}{81} & 0.85 & 0.77 & 0.81 & \multicolumn{1}{c|}{73} & 0.81 & 0.77 & 0.78 & \multicolumn{1}{c|}{77} & 0.80 & 0.81 & \textbf{0.81} & \multicolumn{1}{c|}{82} & 0.79 & 0.83 & 0.81 & 85 \\ \midrule
\multicolumn{1}{l|}{VM} & \multicolumn{1}{c|}{3} & 0.00 & 0.00 & 0.00 & \multicolumn{1}{c|}{1} & 0.50 & 0.33 & 0.40 & \multicolumn{1}{c|}{2} & 0.50 & 0.33 & 0.40 & \multicolumn{1}{c|}{2} & 0.50 & 0.67 & \textbf{0.57} & 4 \\ \midrule
\multicolumn{1}{r}{\textbf{Accuracy}} & \multicolumn{1}{l}{} & \multicolumn{4}{c}{\textbf{0.85}} & \multicolumn{4}{c}{\textbf{0.85}} & \multicolumn{4}{c}{\textbf{0.85}} & \multicolumn{4}{c}{\textbf{0.86}} \\ \bottomrule
\end{tabular}
}
\end{table*}
\begin{table*}[]
\centering
\caption{Per-annotation and Overall Accuracy of the Independent Taggers on Plain Dataset - N/A = annotation not supported by tagger}
\label{tab:plaintable_external}
\resizebox{.8\textwidth}{!}{%

\begin{tabular}{@{}lccccccccccccc@{}}
\toprule
 & \multicolumn{1}{l}{} & \multicolumn{4}{c}{\textbf{SWUM}} & \multicolumn{4}{c}{\textbf{POSSE}} & \multicolumn{4}{c}{\textbf{Stanford}} \\ \midrule
\multicolumn{1}{l|}{Annotation} & \multicolumn{1}{l|}{Total} & Precision & Recall & F1 & \multicolumn{1}{c|}{Total} & Precision & Recall & F1 & \multicolumn{1}{c|}{Total} & Precision & Recall & F1 & Total \\ \midrule
\multicolumn{1}{l|}{CJ} & \multicolumn{1}{c|}{1} & N/A & N/A & N/A & \multicolumn{1}{c|}{N/A} & N/A & N/A & N/A & \multicolumn{1}{c|}{N/A} & \textbf{1.00} & \textbf{1.00} & \textbf{1.00} & 1 \\ \midrule
\multicolumn{1}{l|}{D} & \multicolumn{1}{c|}{7} & 0.33 & 0.14 & 0.20 & \multicolumn{1}{c|}{3} & N/A & N/A & N/A & \multicolumn{1}{c|}{N/A} & \textbf{0.78} & \textbf{1.00} & \textbf{0.88} & 9 \\ \midrule
\multicolumn{1}{l|}{DT} & \multicolumn{1}{c|}{5} & 0.50 & \textbf{0.60} & 0.55 & \multicolumn{1}{c|}{6} & N/A & N/A & N/A & \multicolumn{1}{c|}{N/A} & \textbf{1.00} & 0.40 & \textbf{0.57} & 2 \\ \midrule
\multicolumn{1}{l|}{N} & \multicolumn{1}{c|}{322} & \textbf{0.74} & 0.89 & \textbf{0.81} & \multicolumn{1}{c|}{384} & 0.42 & 0.90 & 0.57 & \multicolumn{1}{c|}{692} & 0.47 & \textbf{0.93} & 0.62 & 647 \\ \midrule
\multicolumn{1}{l|}{NM} & \multicolumn{1}{c|}{415} & 0.78 & \textbf{0.94} & \textbf{0.85} & \multicolumn{1}{c|}{500} & 0.82 & 0.21 & 0.33 & \multicolumn{1}{c|}{106} & \textbf{0.83} & 0.09 & 0.17 & 47 \\ \midrule
\multicolumn{1}{l|}{NPL} & \multicolumn{1}{c|}{78} & N/A & N/A & N/A & \multicolumn{1}{c|}{N/A} & N/A & N/A & N/A & \multicolumn{1}{c|}{N/A} & \textbf{0.86} & \textbf{0.73} & \textbf{0.79} & 66 \\ \midrule
\multicolumn{1}{l|}{P} & \multicolumn{1}{c|}{32} & \textbf{0.88} & 0.44 & 0.58 & \multicolumn{1}{c|}{16} & 0.61 & 0.63 & 0.62 & \multicolumn{1}{c|}{33} & 0.58 & \textbf{0.91} & \textbf{0.71} & 50 \\ \midrule
\multicolumn{1}{l|}{PRE} & \multicolumn{1}{c|}{33} & \textbf{0.50} & \textbf{0.03} & \textbf{0.06} & \multicolumn{1}{c|}{2} & N/A & N/A & N/A & \multicolumn{1}{c|}{N/A} & N/A & N/A & N/A & N/A \\ \midrule
\multicolumn{1}{l|}{V} & \multicolumn{1}{c|}{81} & \textbf{0.89} & 0.72 & \textbf{0.79} & \multicolumn{1}{c|}{65} & 0.77 & 0.75 & 0.76 & \multicolumn{1}{c|}{79} & 0.51 & \textbf{0.90} & 0.65 & 44 \\ \midrule
\multicolumn{1}{l|}{VM} & \multicolumn{1}{c|}{3} & N/A & N/A & N/A & \multicolumn{1}{c|}{N/A} & N/A & N/A & N/A & \multicolumn{1}{c|}{N/A} & \textbf{0.30} & \textbf{1.00} & \textbf{0.46} & 10 \\ \midrule
\multicolumn{1}{r}{\textbf{Accuracy}} & \multicolumn{1}{l}{} & \multicolumn{4}{c}{\textbf{0.77}} & \multicolumn{4}{c}{\textbf{0.47}} & \multicolumn{4}{c}{\textbf{0.52}} \\ \bottomrule
\end{tabular}%
}
\end{table*}

\begin{table}[]
\centering
\caption{Accuracy of independent taggers and best two ensemble taggers in different contexts at the word-level}
\label{tab:word_level_context_accuracy}
\begin{tabular}{@{}l|l|ccccc@{}}
\toprule
 & Total & DTCP & RFCP & SWUM & POSSE & Stanford \\ \midrule
Attribute & 194 & 0.82 & \textbf{0.83} & 0.72 & 0.42 & 0.45 \\ \midrule
Class & 200 & 0.87 & \textbf{0.89} & 0.84 & 0.43 & 0.40 \\ \midrule
Declaration & 184 & 0.83 & \textbf{0.84} & 0.79 & 0.48 & 0.45 \\ \midrule
Function & 231 & \textbf{0.85} & \textbf{0.85} & 0.74 & 0.55 & 0.75 \\ \midrule
Parameter & 168 & \textbf{0.90} & 0.89 & 0.77 & 0.45 & 0.53 \\ \midrule
\multicolumn{1}{r|}{Overall} & 977 & 0.86 & 0.86 & 0.77 & 0.47 & 0.52 \\ \bottomrule
\end{tabular}%
\end{table}

\begin{table*}[]
\centering
\caption{Average Accuracy at the identifier-level}
\label{tab:identifier_level_accuracy}
\begin{tabular}{@{}l|c|cccccccc@{}}
\toprule
\textbf{Category} & \multicolumn{1}{l|}{\textbf{Total}} & \multicolumn{1}{l}{\textbf{DTNA}} & \multicolumn{1}{l}{\textbf{RFNA}} & \multicolumn{1}{l}{\textbf{DTCA}} & \multicolumn{1}{l}{\textbf{RFCA}} & \multicolumn{1}{l}{\textbf{DTNP}} & \multicolumn{1}{l}{\textbf{RFNP}} & \multicolumn{1}{l}{\textbf{DTCP}} & \multicolumn{1}{l}{\textbf{RFCP}} \\ \midrule
Attribute & 76 & 0.70 & 0.70 & \textbf{0.72} & \textbf{0.72} & 0.68 & 0.70 & \textbf{0.72} & \textbf{0.72} \\ \midrule
Class & 77 & 0.78 & 0.78 & 0.77 & \textbf{0.82} & 0.77 & 0.78 & 0.77 & \textbf{0.82} \\ \midrule
Declaration & 77 & 0.66 & 0.71 & 0.66 & 0.73 & \textbf{0.74} & 0.71 & 0.69 & 0.71 \\ \midrule
Function & 77 & 0.62 & 0.62 & 0.68 & 0.65 & 0.64 & 0.64 & \textbf{0.71} & 0.69 \\ \midrule
Parameter & 77 & 0.79 & 0.79 & 0.78 & 0.79 & \textbf{0.81} & \textbf{0.81} & \textbf{0.81} & 0.79 \\ \midrule
\multicolumn{1}{r|}{\textbf{Overall}} & \textbf{384} & 0.71 & 0.72 & 0.72 & 0.74 & 0.73 & 0.73 & 0.74 & \textbf{0.75} \\ \bottomrule
\end{tabular}%
\end{table*}
\begin{table}[]
\centering
\caption{Average accuracy at the identifier-level for state-of-the-art POS taggers}
\label{tab:stateofartaccuracy}
\begin{tabular}{@{}llccc@{}}
\toprule
\textbf{Category} & \textbf{Total} & \multicolumn{1}{l}{\textbf{SWUM}} & \multicolumn{1}{l}{\textbf{POSSE}} & \multicolumn{1}{l}{\textbf{Stanford}} \\ \midrule
\multicolumn{1}{l|}{Attribute} & \multicolumn{1}{l|}{76} & 0.54 & 0.17 & 0.17 \\ \midrule
\multicolumn{1}{l|}{Class} & \multicolumn{1}{l|}{77} & 0.61 & 0.13 & 0.19 \\ \midrule
\multicolumn{1}{l|}{Declaration} & \multicolumn{1}{l|}{77} & 0.65 & 0.27 & 0.22 \\ \midrule
\multicolumn{1}{l|}{Function} & \multicolumn{1}{l|}{77} & 0.48 & 0.30 & 0.29 \\ \midrule
\multicolumn{1}{l|}{Parameter} & \multicolumn{1}{l|}{77} & 0.61 & 0.14 & 0.32 \\ \midrule
\multicolumn{1}{r}{\textbf{Overall}} & \textbf{384} & 0.58 & 0.20 & 0.24 \\ \bottomrule
\end{tabular}%
\end{table}

% \begin{table}[]
% \centering
% \caption{Tukey HSD test results for three independent taggers and top two ensemble taggers at the identifier-level. Bolded values are significant.}
% \label{tab:tukey_hsd_identifier}
% \textcolor{red}{
% \begin{tabular}{@{}l|cc@{}}
% \toprule
% \multicolumn{1}{c|}{Pair under test} & P-Value & Result \\ \midrule
% DTCP vs RFCP & 0.899995 & insignificant \\ \midrule
% DTCP vs SWUM & \textbf{0.004202} & \textbf{p\textless{}0.01} \\ \midrule
% DTCP vs POSSE & \textbf{0.001005} & \textbf{p\textless{}0.01} \\ \midrule
% DTCP vs Stanford & \textbf{0.001005} & \textbf{p\textless{}0.01} \\ \midrule
% RFCP vs SWUM & \textbf{0.002681} & \textbf{p\textless{}0.01} \\ \midrule
% RFCP vs POSSE & \textbf{0.001005} & \textbf{p\textless{}0.01} \\ \midrule
% RFCP vs Stanford & \textbf{0.001005} & \textbf{p\textless{}0.01} \\ \midrule
% SWUM vs POSSE & \textbf{0.001005} & \textbf{p\textless{}0.01} \\ \midrule
% SWUM vs Stanford & \textbf{0.001005} & \textbf{p\textless{}0.01} \\ \midrule
% POSSE vs Stanford & 0.730702 & insignificant \\ \bottomrule
% \end{tabular}%
% }
% \end{table}

\section{Evaluation}\label{evaluation}
\subsection{RQ1: \RQA}
We evaluate accuracy in two ways. The first way is by running 5-fold cross validation using the manually-curated set of 1,335 identifiers. The second way is by running our model on an unseen test set of 384 identifiers. We will split our discussion of RQ1 results into these individual evaluations.

\subsubsection{5-fold test results}
The results from this evaluation are found in Table \ref{tab:five_fold_results_other} and Table \ref{tab:five_fold_results_plain}. These tables give the 5-fold results for five different metrics: accuracy, balanced accuracy, weighted f1, weighted precision, and weighted recall. We ran this 5-fold evaluation on 8 different configurations of our ensemble tagger. You can find the meanings of the abbreviations in Section \ref{setup}. Overall, our results indicate that Random Forest gives the best results regardless of configuration; achieving the highest average in all five metrics used to gauge the quality of our ensemble when compared to decision tree. There is one exception, which is DTCP vs RFCP in Table \ref{tab:five_fold_results_plain}. DTCP achieves the same averages compared with RFCP except with respect to balanced accuracy and weighted f1, where DTCP has better balanced accuracy and RFCP has a better weighted f1. If we look across both tables, then the best configurations that maximize all quality metrics are: DTCP, RFCP, RFNP, and RFCA. Out of those configurations, the authors would advise that the best configuration is DTCP or RFCP. The reason for this is that these configurations 1) are the most accurate on average; 2) use Stanford's conjugations instead of normalizing them away, meaning that they require less dataset preparation; and 3) they use the plain dataset, meaning that they are operating on the full tagset instead of grouping low-frequency annotations together under the OTHER category. This allows them to provide these annotations when they are used to tag identifiers.

\subsubsection{Unseen test set results}
The results from this evaluation are found in Table \ref{tab:othertable} and Table \ref{tab:plaintable}. Each table shows the precision, recall, f1, and accuracy of each ensemble configuration on each individual annotation supported by the ensemble. Table \ref{tab:othertable} has fewer annotations since less-frequent annotations are grouped under the OTHER category. Whereas Table \ref{tab:plaintable} contains all annotations supported in our tagset even if they were very infrequent in the dataset. In addition, these tables show the overall accuracy for each ensemble tagger configuration on the unseen test set. This accuracy is obtained by measuring how many words the ensemble tagger annotated correctly when compared to the manual annotations. One thing to note about these results is that the \textit{total} column on the leftmost side of each table is the total of each annotations in the manually-annotated (i.e., gold) set. Therefore, each configuration would have a different distribution since they each incorrectly annotated some words.

The overall results agree with our 5-fold results. That is, the best taggers tend to be random forest based ensembles. Again, with the exception of DTCP, which is competitive with the other random forest ensembles. RFCP does marginally better than DTCP on the unseen test set, achieving an accuracy of .86 versus DTCP's .85. In addition, since they are trained on the plain dataset, they could be used to annotate tags that are less frequent in production code: DT, CJ, VM, and D; each of which were grouped into the OTHER category in the dataset that RFCA was trained on. It is also notable that the most accurate ensemble configurations were trained on the plain dataset, indicating that the greater tag granularity helped improve the ensemble's output. We come to the same conclusion in this analysis as we did in the 5-fold analysis; DTCP and RFCP are the best ensemble for all of the same advantages explained above alongside retaining the best average values on our metrics. However, RFCP is marginally the better of the two based on our results. One of the primary reasons we still include DTCP as a competitive alternative to RFCP is that DTCP is faster and would scale better for large datasets.

\subsubsection{Comparison with independent taggers} Table \ref{tab:plaintable_external} shows the accuracy of the independent taggers at the word-level. Comparing the data in this table to Table \ref{tab:plaintable}, both DTCP and RFCP outperform or match the other taggers in every individual category. In addition, from Table \ref{tab:word_level_context_accuracy}, we see that DTCP and RFCP maintain their performance advantage at the context level as well.

\subsubsection{Discussion of Feature Importance} In prior work \cite{Newman2020}, we noticed that the individual taggers had strengths and weaknesses that complemented one another. Specifically, Stanford was able to annotate Conjunctions, Digits, Determiners, Noun Plurals, Prepositions, and Verb Modifiers with high accuracy. Meanwhile, SWUM and POSSE tended to outperform Stanford in annotating Noun Modifiers and Verbs. Thus, Stanford's higher balanced accuracy makes perfect sense; it is very complementary to SWUM and POSSE. In addition, we also noticed that word position is important to annotating certain tags, such as Noun Modifiers. This is because, in a noun phrase, the leftmost words tend to be Noun Modifiers while the right-most word is a Noun (i.e., specifically, a head-noun). Another example is that Verbs tend to be in the first position in a function name.

Providing the normalized position helps the ensemble learn these patterns. Normalized position tends to be more effective at this than plain position because normalized position identifies the beginning, middle, and end of an identifier specifically. In contrast, raw position confuses the ensemble since identifiers can be of varying length, making it difficult to identify where the middle and end of an identifier are. Context is, surprisingly, not as important as normalized position. However, it is still part of the best feature set, meaning that it performs better than the subset of features that excludes context but includes normalized position. Thus, knowing whether an identifier is a function name, parameter, etc is still important for annotating with part-of-speech using our approach.

In summary, we have used two different approaches to evaluate our ensemble tagger. In addition, each of these approaches evaluated the ensemble using a set of unseen data to ensure that the ensemble is as general as possible; that the results from its evaluation will translate well to other unseen situations. Based on our data, DTCP and RFCP are equivalent in terms of average performance and have some minor differences between them when we look at which annotations they are most effective on. There is one advantage that DTCP has over RFCP that may be worth mentioning: it is faster. Since decision trees tend to be simpler models than random forests, DTCP generally annotates more rapidly than RFCP. 

\subsection{RQ2: \RQB}
In RQ1, we explore the accuracy of our ensemble tagger at the level of individual words. That is, we want to know how many words it correctly annotates in our dataset. However, the number of correctly annotated words does not give us the full picture. Since most identifiers in the code are made up of multiple words, it is also important to understand how accurate our ensemble tagger is on full identifiers. For this reason, we took the unseen test set and analyzed how effective our ensemble was on full identifier names.

The results of this analysis are given in Table \ref{tab:identifier_level_accuracy} and Table \ref{tab:stateofartaccuracy}. Table \ref{tab:identifier_level_accuracy} shows the accuracy of our individual ensemble configurations, broken down by the five categories we used in our training set: attribute, class, declaration, function, and parameter. In addition, it shows the total number of each type of identifier in the dataset. Table \ref{tab:stateofartaccuracy} shows the individual accuracy of the three state-of-the-art part-of-speech taggers used to construct our ensemble. The numbers here are lower than in the prior tables from RQ1 because we are measuring full identifier names; if even a single word in the identifier name is mis-annotated, then we consider it incorrect.

Our results show that the overall accuracy of our ensemble on full identifier names is unsurprisingly lower than on individual words; about 11-13\% lower in general. However, the ensemble still performs better than its closest competition according to both prior work \cite{Newman2020} and our own analysis shown in Table \ref{tab:stateofartaccuracy}, where we show the accuracy of individual part-of-speech taggers on the same unseen test set on which we ran our ensemble. Comparing the performance of the ensemble and the individual part-of-speech taggers, we can see that the best configuration of our ensemble (i.e., RFCP) outperforms the best tagger, SWUM, by around +17\% on average while it outperforms POSSE and Stanford by 55\% and 51\% respectively. 

In summary, these results are promising, but not surprising. Our ensemble is trained using the output of these approaches, so we would expect that it is able to learn their mistakes and produce output at a higher accuracy than its constituent taggers. We have shown that we can use an ensemble approach to improve upon part-of-speech tagging approaches on source code identifiers at both the individual word level and the full identifier level.

\begin{table*}[]
\centering
\caption{Top 5 most frequently mis-annotated grammar patterns for each ensemble configuration}
\label{tab:topfivemisannotated}
\begin{tabular}{@{}lccc|lccc@{}}
\toprule
\multicolumn{4}{c}{\textbf{DTCA}} & \multicolumn{4}{c}{\textbf{DTCP}} \\ \midrule
\textbf{Grammar Pattern} & \multicolumn{1}{l}{\textbf{\# Incorrect}} & \multicolumn{1}{l}{\textbf{Actual}} & \multicolumn{1}{l|}{\textbf{Proportion}} & \textbf{Grammar Pattern} & \multicolumn{1}{l}{\textbf{\# Incorrect}} & \multicolumn{1}{l}{\textbf{Actual}} & \multicolumn{1}{l}{\textbf{Proportion}} \\ \midrule
NM NM NM NM N & 6 & 8 & 0.75 & NM NM NM NM N & 6 & 8 & 0.75 \\ \midrule
NM NM NM N & 24 & 38 & 0.63 & NM NM NM N & 25 & 38 & 0.66 \\ \midrule
PRE N & 5 & 9 & 0.56 & PRE N & 5 & 9 & 0.56 \\ \midrule
V NM NM N & 9 & 22 & 0.41 & V NM NM N & 9 & 22 & 0.41 \\ \midrule
V NM NPL & 4 & 12 & 0.33 & NM & 2 & 6 & 0.33 \\ \midrule
\multicolumn{4}{c}{\textbf{RFCA}} & \multicolumn{4}{c}{\textbf{RFCP}} \\ \midrule
\textbf{Grammar Pattern} & \multicolumn{1}{l}{\textbf{\# Incorrect}} & \multicolumn{1}{l}{\textbf{Actual}} & \multicolumn{1}{l|}{\textbf{Proportion}} & \textbf{Grammar Pattern} & \multicolumn{1}{l}{\textbf{\# Incorrect}} & \multicolumn{1}{l}{\textbf{Actual}} & \multicolumn{1}{l}{\textbf{Proportion}} \\ \midrule
PRE N & 6 & 9 & 0.67 & PRE N & 6 & 9 & 0.67 \\ \midrule
NM NM NM NM N & 5 & 8 & 0.63 & NM NM NM N & 21 & 38 & 0.55 \\ \midrule
NM NM NM N & 20 & 38 & 0.53 & NM NM NM NM N & 4 & 8 & 0.50 \\ \midrule
NM NM NM NPL & 2 & 4 & 0.50 & V NM NM N & 9 & 22 & 0.41 \\ \midrule
V   NM NM N & 9 & 22 & 0.41 & PRE NM N & 4 & 12 & 0.33 \\ \midrule
\multicolumn{4}{c}{\textbf{DTNA}} & \multicolumn{4}{c}{\textbf{DTNP}} \\ \midrule
\textbf{Grammar Pattern} & \multicolumn{1}{l}{\textbf{\# Incorrect}} & \multicolumn{1}{l}{\textbf{Actual}} & \multicolumn{1}{l|}{\textbf{Proportion}} & \textbf{Grammar Pattern} & \multicolumn{1}{l}{\textbf{\# Incorrect}} & \multicolumn{1}{l}{\textbf{Actual}} & \multicolumn{1}{l}{\textbf{Proportion}} \\ \midrule
NM NM NM NM N & 7 & 8 & 0.88 & NM NM NM NM N & 7 & 8 & 0.88 \\ \midrule
NM NM NM N & 29 & 38 & 0.76 & NM NM NM N & 31 & 38 & 0.82 \\ \midrule
PRE N & 6 & 9 & 0.67 & PRE N & 5 & 9 & 0.56 \\ \midrule
V NM NM N & 10 & 22 & 0.45 & V NM NM N & 9 & 22 & 0.41 \\ \midrule
PRE NM N & 4 & 12 & 0.33 & V N P N & 2 & 5 & 0.40 \\ \midrule
\multicolumn{4}{c}{\textbf{RFNA}} & \multicolumn{4}{c}{\textbf{RFNP}} \\ \midrule
\textbf{Grammar Pattern} & \multicolumn{1}{l}{\textbf{\# Incorrect}} & \multicolumn{1}{l}{\textbf{Actual}} & \multicolumn{1}{l|}{\textbf{Proportion}} & \textbf{Grammar Pattern} & \multicolumn{1}{l}{\textbf{\# Incorrect}} & \multicolumn{1}{l}{\textbf{Actual}} & \multicolumn{1}{l}{\textbf{Proportion}} \\ \midrule
NM NM NM NM N & 6 & 8 & 0.75 & NM NM NM NM N & 7 & 8 & 0.88 \\ \midrule
NM NM NM N & 28 & 38 & 0.74 & NM NM NM N & 29 & 38 & 0.76 \\ \midrule
PRE N & 6 & 9 & 0.67 & PRE N & 6 & 9 & 0.67 \\ \midrule
V NM NM N & 9 & 22 & 0.41 & V NM NM N & 9 & 22 & 0.41 \\ \midrule
PRE NM N & 4 & 12 & 0.33 & PRE NM N & 4 & 12 & 0.33 \\ \bottomrule
\end{tabular}%
\end{table*}
\subsection{RQ3: \RQC}
We have shown the effectiveness of the ensemble tagger at the level of both individual words and full identifier names. In this research question, we are interested in understanding the \textit{weaknesses} of our ensemble; where can it be improved in future work and what types of identifiers is it more likely to get wrong? To answer this question, we calculated the patterns that were most frequently mis-annotated by our ensemble along with the frequency of these patterns in our data. We then divided the number of mis-annotations by the pattern frequency to get the proportion and sorted from largest to smallest. Table \ref{tab:topfivemisannotated} shows the top five mis-annotated grammar patterns per ensemble configuration along with the frequency and proportion information discussed above. 

The data in this table shows some consistently mis-annotated patterns. In particular, \textit{NM NM NM+ N}, \textit{PRE NM* N}, and \textit{V NM NM N} are all in the top 5 for each ensemble configuration. Where '+' means "one or more" of the annotation to its left and '*' means "zero or more" of the annotation to its left. A high frequency of mis-annotating \textit{PRE NM* N} is unsurprising due to the fact that most ensemble configurations had trouble with annotating PRE; the best ensemble configuration achieving only .043 F1. Note that this low F1 score means that it \textbf{both} mis-annotates some NM+ N patterns by annotating PRE where it should have annotated NM as well as mis-annotating PRE NM+ N patterns as NM+ N; not recognizing the first word as a PRE and instead annotating NM. This fact helps explain one of the other patterns it frequently mis-annotates-- the elongated noun-phrase patterns (\textit{NM NM NM+ N}), since it is typically mis-annotating the leftmost NM as PRE. The other pattern it gets wrong frequently is the elongated verb phrase pattern. The ensemble seems to get shorter verb phrase patterns correct but the longer they become the harder it becomes for the ensemble to annotate them. One reason for this is likely the lack of verb phrases that are greater than four and five words in the training set. This may also have something to do with the fact that we use \textbf{normalized position} as one of our features, as discussed in Section\ref{prepandfeatures}. This feature normalizes the length of an identifier by considering words to one be in one of three places: the beginning, the middle, or the end. This helps recognize the fact that the first and last words in an identifier are more likely to be a specific annotation (e.g., the last noun in an identifier is usually a head-noun, whereas middle-nouns are typically noun-adjuncts).

We manually looked at examples of each of these commonly mis-annotated patterns to understand the characteristics of these types of identifiers that the ensemble finds confusing. To make this analysis simpler, we will focus on the best ensemble configurations: DTCP and RFCP. Our manual analysis of the data shows that the most confusing factor in most of these patterns for both RFCP and DTCP is \textit{PRE}. That is, when these are mis-annotated, it is because the correct annotation contains a preamble. For example, \textit{eglewAndroidFrameBufferTarget} and \textit{mRemoveUserDataResponseArgs} both have a grammar pattern that begins with a Preamble but they are mis-annotated as \textit{NM NM NM NM+ N}.
The one exception is the \textit{V NM NM N} pattern, which tended to be mis-annotated because the correct pattern does not follow a standard verb-phrase pattern. These are function names like \textit{clampFixMaxcolor} (\textit{fix} is an abbreviation for \textit{fixpoint}) and \textit{ActionViewShowMasterPages}, which have non-standard function naming structure; \textit{V N NM N} and \textit{NM N V NM NPL} respectively. This does follow results from prior work \cite{Newman2020}, as we found that functions have the largest number of unique grammar patterns; many function names may follow a non-standard format, and the further they are from a standard verb phrase, the more difficult it may be for our tagger to annotate it correctly. If the reader is interested in common mis-annotations made by POS taggers, please refer to prior work for more information on the types of mistakes they make frequently \cite{Newman2020}.

\subsubsection{Discussion of Feature Importance} The results to this RQ show us that more context, in the form of features, is likely required to increase the accuracy of the ensemble. Specifically, context that can 1) help identify Preambles, such as word frequency or system naming conventions. And 2) identify stereotype-like \cite{Dragan:2006} information that could tell the ensemble when it might see a function name that does not follow verb phrase patterns. Of course, other types of context may also be helpful, but these are two types of context that, based on our observations, would highly-likely increase the accuracy of our ensemble.

In summary, our ensemble has more trouble with longer identifier names; particularly longer verb-phrase identifiers in general and longer noun-phrase identifiers which contain a preamble. We found that the mis-annotated verb phrase identifiers are typically function names that do not follow a standard verb-phrase structure, while the mis-annotated noun phrases tended to be elongated and contain a preamble. By far, the most confounding factor for our ensemble is when an identifier contains a preamble. Improved Preamble detection is possible, and we plan to address it in the near future, but it requires an analysis of the system-to-be annotated before commencing annotations. Specifically, to detect preambles, we have to detect frequently-occurring identifier prefixes and then determine whether those prefixes follow the preamble rules specified in Section \ref{grammarpatterndef}.

\section{Threats to validity}\label{threats}
We train our approach on a collection of 1,335 identifiers manually annotated by three of the authors. There is a threat that the annotated set contains imperfections. To mitigate this, we used cross-validation, where all annotators performed a validation step on all grammar patterns; each grammar pattern was validated by two annotators beside the original annotator. Additionally, the dataset is publicly available; future corrections are possible. We calculated that a statistically representative sample for the size of our dataset of 15 systems is 267 given a 95 confidence level and a confidence interval of 6. We picked 95 and 6 as a trade-off between representativeness of the sample, the amount of manual labor required of the annotators, and the sample size used in prior studies.

The dataset that the ensemble is trained on does not control for design-level decisions that may influence naming structure such as design or architectural patterns used by the systems in our dataset. In addition, the dataset does not include test identifiers. This means that the tagger may not be as effective on test identifiers. In addition, while our 5-fold validation indicates that our ensemble works well on a general set of identifier names, identifiers that are written to specific naming conventions due to project-specific, or domain-specific requirements may confuse the ensemble and cause it to perform less-effectively than our 5-fold analysis indicates.

Related to the prior paragraph, we do not address the issue of annotating programming-language-specific naming patterns in this paper. However, our prior work \cite{Newman2020} provides some insight into this problem. In that paper, we were unable to find a significant difference in the grammar patterns present in C/C++ and Java systems. This does not mean that there are not any, as stated in the same paper \cite{Newman2020}, but that our dataset and the variables we controlled for in constructing the dataset did not show a difference. However, we did notice that C/C++ systems tend to have more abbreviations in their identifiers than Java systems, which can, and did \cite{Newman2020}, hinder the effectiveness of part-of-speech tagger performance. Thus, effectiveness of a part-of-speech tagger can vary in different programming languages due to certain identifier name characteristics, such as the presence of abbreviations. We cannot draw conclusions, but future work should more thoroughly investigate the influence of programming language and other system characteristics on naming patterns and part-of-speech tagging.

Overfitting is a threat for the ensemble itself. We mitigate the threat of overfitting by evaluating using 5-fold cross validation, which creates sets of train-test folds to help validate the generality of approaches like our ensemble tagger. In addition, we constructed a manually annotated testing set of identifiers from systems that are not in the training set. Our ensemble's performance is consistent on unseen data from unseen systems, therefore the risk of overfitting is low. 

We did consider other machine learning approaches. Specifically, we considered Support Vector Classification, Logistic Regression, K-Nearest Neighbors, and Multinomial Naive Bayes alongside Random Forest and Decision Trees. We performed a preliminary evaluation using 5-fold cross validation and hyper-tuning. Decision Trees and Random Forest outperformed these other approaches in terms of F1, Accuracy, and Balanced accuracy. However, our preliminary analysis should not be considered the final word on whether these other algorithms could be used to perform POS tagging accurately, because there may be other configurations or features which we have not considered in this paper. To help mitigate the threat that there are other/better ways to create and apply the ensemble, we have made our data and the ensemble tagger publicly available for both use and further research (see Section \ref{methodology}). It is worth noting that our purpose in this paper is not to prove the indisputably \textbf{best} method and feature set for creating an ensemble tagger, but rather to create a useful tool for the research community that outperforms what is currently available and supports future research to improve the state-of-the art both in terms of POS tagging and our understanding of identifier naming practices and patterns. We will continue incrementally improving the ensemble through further research outlined in the Discussion section.

% \marcos{I'm generally not a fan of related work sections in the end of the paper. I think it makes more sense to have it after the introduction to set the stage to the main content of the paper. Maybe having the RW in in the end of the paper is more common in SE and in this case, no problem.}

\section{Discussion \& Conclusion} \label{Discussion+conclusion}
In this paper, we have described and evaluated an ensemble tagging technique which combines the output of three state-of-the-art part-of-speech taggers using the random forest and decision tree machine learning algorithms. We have shown that our approach improves the output of these taggers by learning their common mistakes and adjusting to correct these mistakes. The evaluation shows that the ensemble tagger achieves an accuracy, at the identifier level, up to 75\%; a +17\% point increase on top of the best runner-up tagger's accuracy (i.e., SWUM at 58\%). In addition, our ensemble approach achieves up to 86\% accuracy at the word-level. 

We use several features to achieve this quality, including the annotations from SWUM, POSSE, and Stanford; the context of the identifier (e.g., whether it is a function or class identifier), and the normalized position of each word within the identifier. These features were compared against four other features: word, data type, position, and max position. We found that the addition of these features degraded the performance of the ensemble, so we did not use them in the final version of our ensemble tagger. While this does not mean that they are bad features to include in future research, it does indicate that they can negatively impact performance for our approach and should be used carefully if they are used at all. 

Our ensemble has several potential weaknesses which must be evaluated and accounted for in future research. These weaknesses are as follows:
\begin{enumerate}
    \item \textbf{Identifiers containing preambles, and non-verb-phrase function names.} Based on RQ3, our approach may have decreased accuracy on identifier names that grow past the average identifier size and are 1) function names that do not follow a standard verb phrase pattern or 2) elongated noun-phrases that contain a preamble.
    \item \textbf{Test identifier names.} Our ensemble is trained on a data set that does not include test identifier names. Therefore, under the supported \cite{Zhang2015ASE, Zhang2015ASE} assumption that test identifiers have a different structure than production identifiers, our approach will be less accurate on these identifiers under its current training set.
    \item \textbf{Design-, or architecture-, specific identifier structures.} In prior work, we noticed that certain identifier naming structures seem to indicate the use of certain design or architectural patterns \cite{Newman2020}. However, we did not control for these in the construction of the dataset that is used to train this ensemble approach and there is no dataset of part-of-speech tagged identifiers that controls for these aspects to our knowledge. It is possible that the accuracy of our approach is negatively influenced due to not taking these design decisions into account. 
\end{enumerate}
Therefore, in future work, we plan to expand our dataset to handle these concerns. This will have the advantage of both increasing the size of our dataset as well as increasing our understanding of how different software development contexts influence identifier names.

In conclusion, we have presented an ensemble part-of-speech tagger for source code identifiers. The ensemble uses several features of identifiers, as well as three state-of-the-art part-of-speech taggers, to improve upon the quality of identifier part-of-speech annotations. In addition, we have released the tool for use by other researchers and it has been integrated with the srcML framework \cite{collard:2016}. We are also investigating ways to remove its reliance on other taggers' output and create a fully self-contained part-of-speech tagger for source code.

\section{ACKNOWLEDGEMENT} \label{sec:acknowledgement}
This material is based upon work supported by the National Science Foundation under Grant No. 1850412.  We would also like to thank Aidan White for initial help with the manual-annotation effort. 
\section{DISCLAIMER}
The views expressed in this paper are not affiliated or endorsed by BNY Mellon in any way.
\balance
\bibliography{references}

\begin{thebibliography}{10}
\providecommand{\url}[1]{#1}
\csname url@samestyle\endcsname
\providecommand{\newblock}{\relax}
\providecommand{\bibinfo}[2]{#2}
\providecommand{\BIBentrySTDinterwordspacing}{\spaceskip=0pt\relax}
\providecommand{\BIBentryALTinterwordstretchfactor}{4}
\providecommand{\BIBentryALTinterwordspacing}{\spaceskip=\fontdimen2\font plus
\BIBentryALTinterwordstretchfactor\fontdimen3\font minus
  \fontdimen4\font\relax}
\providecommand{\BIBforeignlanguage}[2]{{%
\expandafter\ifx\csname l@#1\endcsname\relax
\typeout{** WARNING: IEEEtran.bst: No hyphenation pattern has been}%
\typeout{** loaded for the language `#1'. Using the pattern for}%
\typeout{** the default language instead.}%
\else
\language=\csname l@#1\endcsname
\fi
#2}}
\providecommand{\BIBdecl}{\relax}
\BIBdecl

\bibitem{Corbi1989}
T.~A. {Corbi}, ``Program understanding: Challenge for the 1990s,'' \emph{IBM
  Systems Journal}, vol.~28, no.~2, pp. 294--306, 1989.

\bibitem{Martin:2008}
R.~C. Martin, \emph{Clean Code: A Handbook of Agile Software Craftsmanship},
  1st~ed.\hskip 1em plus 0.5em minus 0.4em\relax Upper Saddle River, NJ, USA:
  Prentice Hall PTR, 2008.

\bibitem{vonMayrhauser:1997}
\BIBentryALTinterwordspacing
A.~von Mayrhauser and A.~M. Vans, ``Program understanding behavior during
  debugging of large scale software,'' in \emph{Papers Presented at the Seventh
  Workshop on Empirical Studies of Programmers}, ser. ESP '97.\hskip 1em plus
  0.5em minus 0.4em\relax New York, NY, USA: ACM, 1997, pp. 157--179. [Online].
  Available: \url{http://doi.acm.org/10.1145/266399.266414}
\BIBentrySTDinterwordspacing

\bibitem{Fisher:2006}
\BIBentryALTinterwordspacing
M.~Fisher, A.~Cox, and L.~Zhao, ``Using sex differences to link spatial
  cognition and program comprehension,'' in \emph{Proceedings of the 22Nd IEEE
  International Conference on Software Maintenance}, ser. ICSM '06.\hskip 1em
  plus 0.5em minus 0.4em\relax Washington, DC, USA: IEEE Computer Society,
  2006, pp. 289--298. [Online]. Available:
  \url{https://doi.org/10.1109/ICSM.2006.72}
\BIBentrySTDinterwordspacing

\bibitem{Deissenbock:2005}
\BIBentryALTinterwordspacing
F.~Deissenboeck and M.~Pizka, ``Concise and consistent naming,'' \emph{Software
  Quality Journal}, vol.~14, no.~3, p. 261–282, Sep. 2006. [Online].
  Available: \url{https://doi.org/10.1007/s11219-006-9219-1}
\BIBentrySTDinterwordspacing

\bibitem{Binkley:2018}
\BIBentryALTinterwordspacing
D.~Binkley, D.~Lawrie, and C.~Morrell, ``The need for software specific natural
  language techniques,'' \emph{Empirical Softw. Engg.}, vol.~23, no.~4, pp.
  2398--2425, Aug. 2018. [Online]. Available:
  \url{https://doi.org/10.1007/s10664-017-9566-5}
\BIBentrySTDinterwordspacing

\bibitem{Jongeling:2017}
R.~Jongeling, P.~Sarkar, S.~Datta, and A.~Serebrenik, ``On negative results
  when using sentiment analysis tools for software engineering research,''
  \emph{Empirical Software Engineering}, 01 2017.

\bibitem{Manning:1999}
C.~D. Manning and H.~Sch\"{u}tze, \emph{Foundations of Statistical Natural
  Language Processing}.\hskip 1em plus 0.5em minus 0.4em\relax Cambridge, MA,
  USA: MIT Press, 1999.

\bibitem{miller1995wordnet}
G.~A. Miller, ``Wordnet: a lexical database for english,'' \emph{Communications
  of the ACM}, vol.~38, no.~11, pp. 39--41, 1995.

\bibitem{Liu:2019}
K.~Liu, D.~Kim, T.~F.~Bissyandé, T.~Kim, K.~Kim, A.~Koyuncu, S.~Kim, and
  Y.~Le~Traon, ``Learning to spot and refactor inconsistent method names,'' in
  \emph{Proceedings of the 40th International Conference on Software
  Engineering}, ser. ICSE 2019.\hskip 1em plus 0.5em minus 0.4em\relax New
  York, NY, USA: ACM, 2019.

\bibitem{Arnaoudova:2014}
\BIBentryALTinterwordspacing
V.~Arnaoudova, L.~M. Eshkevari, M.~D. Penta, R.~Oliveto, G.~Antoniol, and Y.-G.
  Gueheneuc, ``Repent: Analyzing the nature of identifier renamings,''
  \emph{IEEE Trans. Softw. Eng.}, vol.~40, no.~5, pp. 502--532, May 2014.
  [Online]. Available: \url{https://doi.org/10.1109/TSE.2014.2312942}
\BIBentrySTDinterwordspacing

\bibitem{Perumascam}
A.~Peruma, M.~W. Mkaouer, M.~J. Decker, and C.~D. Newman, ``Contextualizing
  rename decisions using refactorings and commit messages,'' in
  \emph{Proceedings of the 19th IEEE International Working Conference on Source
  Code Analysis and Manipulation}.\hskip 1em plus 0.5em minus 0.4em\relax IEEE,
  2019.

\bibitem{PERUMA2020110704}
A.~Peruma, M.~W. Mkaouer, M.~Decker, and C.~Newman, ``Contextualizing rename
  decisions using refactorings, commit messages, and data types,''
  \emph{Journal of Systems and Software}, vol. 169, p. 110704, 06 2020.

\bibitem{Arnaoudova:2013}
V.~{Arnaoudova}, M.~{Di Penta}, G.~{Antoniol}, and Y.~{Guéhéneuc}, ``A new
  family of software anti-patterns: Linguistic anti-patterns,'' in \emph{2013
  17th European Conference on Software Maintenance and Reengineering}, March
  2013, pp. 187--196.

\bibitem{Hill:2014}
\BIBentryALTinterwordspacing
E.~Hill, D.~Binkley, D.~Lawrie, L.~Pollock, and K.~Vijay-Shanker, ``An
  empirical study of identifier splitting techniques,'' \emph{Empirical Softw.
  Engg.}, vol.~19, no.~6, pp. 1754--1780, Dec. 2014. [Online]. Available:
  \url{http://dx.doi.org/10.1007/s10664-013-9261-0}
\BIBentrySTDinterwordspacing

\bibitem{HillSWUM:2010}
E.~Hill, ``Integrating natural language and program structure information to
  improve software search and exploration,'' Ph.D. dissertation, Newark, DE,
  USA, 2010, aAI3423409.

\bibitem{Gupta:2013}
S.~{Gupta}, S.~{Malik}, L.~{Pollock}, and K.~{Vijay-Shanker}, ``Part-of-speech
  tagging of program identifiers for improved text-based software engineering
  tools,'' in \emph{2013 21st International Conference on Program Comprehension
  (ICPC)}, May 2013, pp. 3--12.

\bibitem{Olney2016}
W.~{Olney}, E.~{Hill}, C.~{Thurber}, and B.~{Lemma}, ``Part of speech tagging
  java method names,'' in \emph{2016 IEEE International Conference on Software
  Maintenance and Evolution (ICSME)}, Oct 2016, pp. 483--487.

\bibitem{Host:2009}
\BIBentryALTinterwordspacing
E.~W. H{\o}st and B.~M. {\O}stvold, ``Debugging method names,'' in
  \emph{Proceedings of the 23rd European Conference on ECOOP 2009 ---
  Object-Oriented Programming}, ser. Genoa.\hskip 1em plus 0.5em minus
  0.4em\relax Berlin, Heidelberg: Springer-Verlag, 2009, pp. 294--317.
  [Online]. Available: \url{http://dx.doi.org/10.1007/978-3-642-03013-0_14}
\BIBentrySTDinterwordspacing

\bibitem{Shepherd:2009}
E.~{Hill}, L.~{Pollock}, and K.~{Vijay-Shanker}, ``Automatically capturing
  source code context of nl-queries for software maintenance and reuse,'' in
  \emph{2009 IEEE 31st International Conference on Software Engineering}, May
  2009, pp. 232--242.

\bibitem{butler:2011}
S.~{Butler}, M.~{Wermelinger}, Y.~{Yu}, and H.~{Sharp}, ``Mining java class
  naming conventions,'' in \emph{2011 27th IEEE International Conference on
  Software Maintenance (ICSM)}, Sep. 2011, pp. 93--102.

\bibitem{butler:2015}
S.~{Butler}, M.~{Wermelinger}, and Y.~{Yu}, ``A survey of the forms of java
  reference names,'' in \emph{2015 IEEE 23rd International Conference on
  Program Comprehension}, May 2015, pp. 196--206.

\bibitem{Liblit06cognitiveperspectives}
B.~Liblit, A.~Begel, and E.~Sweetser, ``Cognitive perspectives on the role of
  naming in computer programs,'' in \emph{In Proc. of the 18th Annual
  Psychology of Programming Workshop}, 2006.

\bibitem{posit:2020}
\BIBentryALTinterwordspacing
P.-P. P\^{a}rtachi, S.~K. Dash, C.~Treude, and E.~T. Barr, ``Posit:
  Simultaneously tagging natural and programming languages,'' in
  \emph{Proceedings of the ACM/IEEE 42nd International Conference on Software
  Engineering}, ser. ICSE '20.\hskip 1em plus 0.5em minus 0.4em\relax New York,
  NY, USA: Association for Computing Machinery, 2020, p. 1348–1358. [Online].
  Available: \url{https://doi.org/10.1145/3377811.3380440}
\BIBentrySTDinterwordspacing

\bibitem{Newman2020}
\BIBentryALTinterwordspacing
C.~D. Newman, R.~S. AlSuhaibani, M.~J. Decker, A.~Peruma, D.~Kaushik, M.~W.
  Mkaouer, and E.~Hill, ``On the generation, structure, and semantics of
  grammar patterns in source code identifiers,'' \emph{Journal of Systems and
  Software}, vol. 170, p. 110740, Dec 2020. [Online]. Available:
  \url{http://dx.doi.org/10.1016/j.jss.2020.110740}
\BIBentrySTDinterwordspacing

\bibitem{decisiontrees}
\BIBentryALTinterwordspacing
D.~H. Moore~II, ``Classification and regression trees, by leo breiman, jerome
  h. friedman, richard a. olshen, and charles j. stone. brooks/cole publishing,
  monterey, 1984,358 pages, \$27.95,'' \emph{Cytometry}, vol.~8, no.~5, pp.
  534--535, 1987. [Online]. Available:
  \url{https://onlinelibrary.wiley.com/doi/abs/10.1002/cyto.990080516}
\BIBentrySTDinterwordspacing

\bibitem{randomforest}
\BIBentryALTinterwordspacing
L.~Breiman, ``Random forests,'' \emph{Mach. Learn.}, vol.~45, no.~1, p. 5–32,
  Oct. 2001. [Online]. Available: \url{https://doi.org/10.1023/A:1010933404324}
\BIBentrySTDinterwordspacing

\bibitem{Roy2020ICPC}
\BIBentryALTinterwordspacing
D.~Roy, S.~Fakhoury, J.~Lee, and V.~Arnaoudova, ``A model to detect readability
  improvements in incremental changes,'' in \emph{Proceedings of the 28th
  International Conference on Program Comprehension}, ser. ICPC '20.\hskip 1em
  plus 0.5em minus 0.4em\relax New York, NY, USA: Association for Computing
  Machinery, 2020, p. 25–36. [Online]. Available:
  \url{https://doi.org/10.1145/3387904.3389255}
\BIBentrySTDinterwordspacing

\bibitem{AlOmar2021ESWA}
\BIBentryALTinterwordspacing
E.~A. AlOmar, A.~Peruma, M.~W. Mkaouer, C.~Newman, A.~Ouni, and M.~Kessentini,
  ``How we refactor and how we document it? on the use of supervised machine
  learning algorithms to classify refactoring documentation,'' \emph{Expert
  Systems with Applications}, p. 114176, 2020. [Online]. Available:
  \url{http://www.sciencedirect.com/science/article/pii/S095741742030912X}
\BIBentrySTDinterwordspacing

\bibitem{DellOrletta2009EnsembleSF}
F.~Dell'Orletta, ``Ensemble system for part-of-speech tagging,'' in
  \emph{Proceedings of EVALITA 2009}, 2009.

\bibitem{2021:ICPC:METHODS}
A.~Peruma, E.~Hu, J.~Chen, E.~A. Alomar, M.~W. Mkaouer, and C.~D. Newman,
  ``Using grammar patterns to interpret test method name evolution,'' in
  \emph{Proceedings of the 29th International Conference on Program
  Comprehension}, ser. ICPC '21.\hskip 1em plus 0.5em minus 0.4em\relax New
  York, NY, USA: Association for Computing Machinery, May 2021.

\bibitem{Toutanova:StanfordTagger}
\BIBentryALTinterwordspacing
K.~Toutanova and C.~D. Manning, ``Enriching the knowledge sources used in a
  maximum entropy part-of-speech tagger,'' in \emph{Proceedings of the 2000
  Joint SIGDAT Conference on Empirical Methods in Natural Language Processing
  and Very Large Corpora: Held in Conjunction with the 38th Annual Meeting of
  the Association for Computational Linguistics - Volume 13}, ser. EMNLP
  '00.\hskip 1em plus 0.5em minus 0.4em\relax Stroudsburg, PA, USA: Association
  for Computational Linguistics, 2000, pp. 63--70. [Online]. Available:
  \url{https://doi.org/10.3115/1117794.1117802}
\BIBentrySTDinterwordspacing

\bibitem{butler2010exploring}
S.~Butler, M.~Wermelinger, Y.~Yu, and H.~Sharp, ``Exploring the influence of
  identifier names on code quality: An empirical study,'' in \emph{Software
  Maintenance and Reengineering (CSMR), 2010 14th European Conference
  on}.\hskip 1em plus 0.5em minus 0.4em\relax IEEE, 2010, pp. 156--165.

\bibitem{hillamap}
\BIBentryALTinterwordspacing
E.~Hill, Z.~P. Fry, H.~Boyd, G.~Sridhara, Y.~Novikova, L.~Pollock, and
  K.~Vijay-Shanker, ``Amap: Automatically mining abbreviation expansions in
  programs to enhance software maintenance tools,'' in \emph{Proceedings of the
  2008 International Working Conference on Mining Software Repositories}, ser.
  MSR ’08.\hskip 1em plus 0.5em minus 0.4em\relax New York, NY, USA:
  Association for Computing Machinery, 2008, p. 79–88. [Online]. Available:
  \url{https://doi.org/10.1145/1370750.1370771}
\BIBentrySTDinterwordspacing

\bibitem{hungariannotation}
C.~Simonyi and M.~Heller, ``The hungarian revolution,'' \emph{BYTE}, vol.~16,
  no.~8, p. 131–ff., Aug. 1991.

\bibitem{Peruma:2018:EIW:3242163.3242169}
\BIBentryALTinterwordspacing
A.~Peruma, M.~W. Mkaouer, M.~J. Decker, and C.~D. Newman, ``An empirical
  investigation of how and why developers rename identifiers,'' in
  \emph{International Workshop on Refactoring 2018}, 2018. [Online]. Available:
  \url{http://doi.acm.org/10.1145/3242163.3242169}
\BIBentrySTDinterwordspacing

\bibitem{Butler:2009}
S.~{Butler}, M.~{Wermelinger}, Y.~{Yu}, and H.~{Sharp}, ``Relating identifier
  naming flaws and code quality: An empirical study,'' in \emph{2009 16th
  Working Conference on Reverse Engineering}, Oct 2009, pp. 31--35.

\bibitem{Binkley:2011}
\BIBentryALTinterwordspacing
D.~Binkley, M.~Hearn, and D.~Lawrie, ``Improving identifier informativeness
  using part of speech information,'' in \emph{Proceedings of the 8th Working
  Conference on Mining Software Repositories}, ser. MSR '11.\hskip 1em plus
  0.5em minus 0.4em\relax New York, NY, USA: ACM, 2011, pp. 203--206. [Online].
  Available: \url{http://doi.acm.org/10.1145/1985441.1985471}
\BIBentrySTDinterwordspacing

\bibitem{Wu2020JSS}
\BIBentryALTinterwordspacing
J.~Wu and J.~Clause, ``A pattern-based approach to detect and improve
  non-descriptive test names,'' \emph{Journal of Systems and Software}, vol.
  168, p. 110639, 2020. [Online]. Available:
  \url{http://www.sciencedirect.com/science/article/pii/S0164121220301126}
\BIBentrySTDinterwordspacing

\bibitem{Nguyen:2020}
\BIBentryALTinterwordspacing
S.~Nguyen, H.~Phan, T.~Le, and T.~N. Nguyen, ``Suggesting natural method names
  to check name consistencies,'' in \emph{Proceedings of the ACM/IEEE 42nd
  International Conference on Software Engineering}, ser. ICSE '20.\hskip 1em
  plus 0.5em minus 0.4em\relax New York, NY, USA: Association for Computing
  Machinery, 2020, p. 1372–1384. [Online]. Available:
  \url{https://doi.org/10.1145/3377811.3380926}
\BIBentrySTDinterwordspacing

\bibitem{tonella:1999}
C.~{Caprile} and P.~{Tonella}, ``Nomen est omen: analyzing the language of
  function identifiers,'' in \emph{Sixth Working Conference on Reverse
  Engineering (Cat. No.PR00303)}, Oct 1999, pp. 112--122.

\bibitem{tonella:2000}
{Caprile} and {Tonella}, ``Restructuring program identifier names,'' in
  \emph{Proceedings 2000 International Conference on Software Maintenance}, Oct
  2000, pp. 97--107.

\bibitem{Shepherd:2007}
\BIBentryALTinterwordspacing
D.~Shepherd, Z.~P. Fry, E.~Hill, L.~Pollock, and K.~Vijay-Shanker, ``Using
  natural language program analysis to locate and understand action-oriented
  concerns,'' in \emph{Proceedings of the 6th International Conference on
  Aspect-oriented Software Development}, ser. AOSD '07.\hskip 1em plus 0.5em
  minus 0.4em\relax New York, NY, USA: ACM, 2007, pp. 212--224. [Online].
  Available: \url{http://doi.acm.org/10.1145/1218563.1218587}
\BIBentrySTDinterwordspacing

\bibitem{fry:2008}
Z.~P. {Fry}, D.~{Shepherd}, E.~{Hill}, L.~{Pollock}, and K.~{Vijay-Shanker},
  ``Analysing source code: looking for useful verb-direct object pairs in all
  the right places,'' \emph{IET Software}, vol.~2, no.~1, pp. 27--36, February
  2008.

\bibitem{singer:2008}
J.~{Singer} and C.~{Kirkham}, ``Exploiting the correspondence between micro
  patterns and class names,'' in \emph{2008 Eighth IEEE International Working
  Conference on Source Code Analysis and Manipulation}, Sep. 2008, pp. 67--76.

\bibitem{hostdissertation}
E.~W. H{\o}st, ``Meaningful method names,'' 2011.

\bibitem{HostLexicon}
E.~Host and B.~Ostvold, ``The programmer's lexicon, volume i: The verbs,'' 09
  2007, pp. 193 -- 202.

\bibitem{hostphrasebook}
E.~W. H{\o}st and B.~M. {\O}stvold, ``The java programmer's phrase book,'' in
  \emph{Software Language Engineering}, D.~Ga{\v{s}}evi{\'{c}}, R.~L{\"a}mmel,
  and E.~Van~Wyk, Eds.\hskip 1em plus 0.5em minus 0.4em\relax Berlin,
  Heidelberg: Springer Berlin Heidelberg, 2009, pp. 322--341.

\bibitem{Hst2011CanonicalMN}
E.~W. "H{\o}st and B.~M. {\O}stvold, ``Canonical method names for java,'' in
  \emph{Software Language Engineering}, B.~"Malloy, S.~Staab, and M.~van~den
  Brand, Eds.\hskip 1em plus 0.5em minus 0.4em\relax Berlin, Heidelberg:
  Springer Berlin Heidelberg, 2011, pp. 226--245.

\bibitem{collard:2016}
M.~L. {Collard} and J.~I. {Maletic}, ``srcml 1.0: Explore, analyze, and
  manipulate source code,'' in \emph{2016 IEEE International Conference on
  Software Maintenance and Evolution (ICSME)}, Oct 2016, pp. 649--649.

\bibitem{Zhang2015ASE}
B.~{Zhang}, E.~{Hill}, and J.~{Clause}, ``Automatically generating test
  templates from test names (n),'' in \emph{2015 30th IEEE/ACM International
  Conference on Automated Software Engineering (ASE)}, 2015, pp. 506--511.

\bibitem{Zhang2016ASE}
\BIBentryALTinterwordspacing
B.~Zhang, E.~Hill, and J.~Clause, ``Towards automatically generating
  descriptive names for unit tests,'' in \emph{Proceedings of the 31st IEEE/ACM
  International Conference on Automated Software Engineering}, ser. ASE
  2016.\hskip 1em plus 0.5em minus 0.4em\relax New York, NY, USA: Association
  for Computing Machinery, 2016, p. 625–636. [Online]. Available:
  \url{https://doi.org/10.1145/2970276.2970342}
\BIBentrySTDinterwordspacing

\bibitem{newmanabbrev}
C.~D. Newman, M.~J. Decker, R.~S. AlSuhaibani, A.~Peruma, D.~Kaushik, and
  E.~Hill, ``An empirical study of abbreviations and expansions in software
  artifacts,'' in \emph{Proceedings of the 35th IEEE International Conference
  on Software Maintenance}.\hskip 1em plus 0.5em minus 0.4em\relax IEEE, 2019.

\bibitem{abebe10}
\BIBentryALTinterwordspacing
S.~L. Abebe and P.~Tonella, ``Natural language parsing of program element names
  for concept extraction,'' in \emph{Proceedings of the 2010 IEEE 18th
  International Conference on Program Comprehension}, ser. ICPC ’10.\hskip
  1em plus 0.5em minus 0.4em\relax USA: IEEE Computer Society, 2010, p.
  156–159. [Online]. Available: \url{https://doi.org/10.1109/ICPC.2010.29}
\BIBentrySTDinterwordspacing

\bibitem{Albon201Machine}
\BIBentryALTinterwordspacing
C.~Albon, \emph{Machine Learning with Python Cookbook: Practical Solutions from
  Preprocessing to Deep Learning}.\hskip 1em plus 0.5em minus 0.4em\relax
  O'Reilly Media, 2018. [Online]. Available:
  \url{https://books.google.com/books?id=kIhQDwAAQBAJ}
\BIBentrySTDinterwordspacing

\bibitem{dangeti2017statistics}
P.~Dangeti, \emph{Statistics for Machine Learning}.\hskip 1em plus 0.5em minus
  0.4em\relax Packt Publishing, 2017.

\bibitem{Dragan:2006}
\BIBentryALTinterwordspacing
N.~Dragan, M.~L. Collard, and J.~I. Maletic, ``Reverse engineering method
  stereotypes,'' in \emph{Proceedings of the 22Nd IEEE International Conference
  on Software Maintenance}, ser. ICSM '06.\hskip 1em plus 0.5em minus
  0.4em\relax Washington, DC, USA: IEEE Computer Society, 2006, pp. 24--34.
  [Online]. Available: \url{http://dx.doi.org/10.1109/ICSM.2006.54}
\BIBentrySTDinterwordspacing

\end{thebibliography}
\end{document}